\documentclass[twocolumn]{aa}

\usepackage{graphicx}
\usepackage{epstopdf}
\usepackage{times,epsfig} 
\usepackage{amssymb}
\usepackage{mathrsfs}  
\usepackage{booktabs}

\usepackage{natbib}
\bibpunct{(}{)}{;}{a}{}{,} 
\usepackage{caption}

\usepackage[switch]{lineno}

\usepackage{amsmath}
\usepackage[english]{babel}
\usepackage{longtable}
\usepackage{afterpage}
\usepackage{url}
\usepackage{hyperref}
\usepackage{xcolor}
\usepackage{soul}
\definecolor{dark-red}{rgb}{0.4,0.15,0.15}
\definecolor{dark-blue}{rgb}{0.15,0.15,0.4}
\definecolor{medium-blue}{rgb}{0,0,0.5}
\hypersetup{
    colorlinks, linkcolor={dark-red},
    citecolor={dark-blue}, urlcolor={medium-blue}
}

\newcommand{\beqa}{\begin{eqnarray}} 
\newcommand{\eeqa}{\end{eqnarray}}

\newcommand{\bsub}{\begin{subequations}}
\newcommand{\esub}{\end{subequations}}
\newcommand{\beal}{\begin{align}}
\newcommand{\ealn}{\end{align}}

\newcommand{\kms}{{\rm \  km~s^{-1}}}
\newcommand{\msun}{M$_{\sun}$}

\newcommand{\Ha}{H$\alpha$}
\newcommand{\Hb}{H$\beta$}

\newcommand{\wll}{\lambda \lambda}
\newcommand{\ccm}{{\rm cm}^{-3}}

\newcommand{\Msun}{{\ensuremath{\mathrm{M}_{\odot}}}}

\begin{document}

\title{SN 2019zrk, a bright SN 2009ip analog with a precursor \thanks{Table 1 is only available in electronic form
at the CDS via anonymous ftp to cdsarc.u-strasbg.fr (130.79.128.5)
or via http://cdsweb.u-strasbg.fr/cgi-bin/qcat?J/A+A/}}

\author{Claes Fransson\inst{1} 
\and 
Jesper Sollerman\inst{1} 
\and
 Nora L. Strotjohann\inst{2} 
\and
Sheng Yang\inst{1}
\and
 Steve Schulze\inst{3} 
\and
Cristina Barbarino\inst{1}
\and
Erik C. Kool\inst{1}
\and
Eran O. Ofek\inst{2}
\and
Arien Crellin-Quick\inst{4}
\and
Kishalay De\inst{5}
\and
Andrew J. Drake\inst{6}
\and
Christoffer Fremling\inst{5}
\and 
Avishay Gal-Yam\inst{7}
\and
Anna Y. Q. Ho\inst{4,8}
\and
Mansi M. Kasliwal\inst{5}
}

\institute{Department of Astronomy, The Oskar Klein Center, Stockholm University, AlbaNova, 10691 Stockholm, Sweden
\and
 Benoziyo Center for Astrophysics, The Weizmann Institute of Science, Rehovot 76100, Israel
\and
 Department of Physics, The Oskar Klein Center, Stockholm University, AlbaNova, 10691 Stockholm, Sweden
\and
 Department of Astronomy, University of California, Berkeley, 501 Campbell Hall, Berkeley, CA, 94720, USA
\and 
Division of Physics, Mathematics and Astronomy, California Institute of
Technology, Pasadena, CA 91125, USA
\and
Cahill Center for Astrophysics, California Institute of Technology, MC 249-17, 1200 E California Boulevard, Pasadena, CA, 91125, USA
\and
Department of Particle Physics and Astrophysics, Weizmann Institute of Science, 76100 Rehovot, Israel
\and
Miller Institute for Basic Research in Science, 468 Donner Lab, Berkeley, CA 94720, USA
}
\date{}
%
\abstract{
We present photometric and spectroscopic observations of the Type IIn supernova SN\,2019zrk (also known as ZTF\,20aacbyec). The SN shows a $\gtrsim$ 100 day precursor, with a slow rise, followed by a rapid rise to $M \approx -19.2$ in the $r$ and $g$ bands.   The post-peak light-curve decline is well fit with an exponential decay with a timescale of $\sim 39$ days, but it shows prominent undulations, with an amplitude of $\sim 1$ mag. Both the light curve and spectra are dominated by an interaction with a dense circumstellar medium (CSM), probably from previous mass ejections. The spectra evolve from a scattering-dominated Type IIn spectrum to a spectrum with strong P-Cygni absorptions. The expansion velocity is high, $\sim 16,000 \kms$, even in the last spectra.  The last spectrum $\sim 110$ days after the main eruption reveals no evidence for advanced nucleosynthesis. 
From analysis of the spectra and light curves, we estimate the mass-loss rate to be $\sim 4 \times 10^{-2} \ \Msun$ \ yr$^{-1}$ for a  CSM velocity of $100 \kms$, and a CSM mass of  $\gtrsim 1 \ \Msun$. We find strong similarities for both the precursor, general light curve, and spectral evolution with SN 2009ip and similar SNe, although SN 2019zrk displays a brighter peak magnitude. 
Different scenarios for the nature of the 09ip-class of SNe, based on pulsational pair instability eruptions, wave heating, and mergers, are discussed.  
}
%
\authorrunning{Fransson, Sollerman, Strotjohann et al.}
\titlerunning{SN 2019zrk}
\maketitle
%
\section{Introduction}
\label{sec:intro}

Core-collapse (CC) supernovae (SNe) are explosions of massive stars ($\gtrsim8~M_\odot$).
The large variety of CC SNe
is determined by the progenitor mass at the time of CC and
by the mass-loss history leading up to the explosion. 
Type IIP SNe are the best understood class when it comes to their progenitor stars, which are mostly red supergiants \citep{Smartt:2009aa} exploding while still retaining a massive H envelope. 
Type IIn SNe, on the other hand, may originate from more than one progenitor channel -- an ambiguity that is often difficult to unveil as the circumstellar matter (CSM) interaction hides the underlying SN ejecta \citep[see][for reviews]{ChevalierFransson2017,Galyam2017,Smith2017}. 
A discriminating factor is the mass of the envelope, which may range from $\ll 1$ \msun \ to several \msun. This is reflected in the luminosity, shock velocity, and the duration of the optically thick phase. An example of a Type IIn SN with a 
low-mass CSM is SN 1998S with only a week-long SN IIn phase \cite[e.g.,][]{2004MNRAS.352..457P}, while SNe 2010jl \citep{fransson10jl,Ofek2014} and 2015da \citep{Tartaglia20} are examples of SNe with a massive CSM, resulting in a month- to year-long SN IIn phase. In the context of this paper, the enigmatic Type IIn SN 2009ip is especially interesting \citep[see][for a review]{Fraser2020}.  The nature of the SN IIn progenitors is, however, still not well understood even though the relation with host metallicity, for example, indicates that the former low-mass CSM SNe IIn are related to red supergiants, while the long-duration Type IIn SNe might be related to luminous blue variables (LBVs) \citep{Taddia2015}.  The SN 2009ip like SNe are even less understood.

In this paper we present SN\,2019zrk, which is clearly a H-rich SN, but which showed an unusual light curve evolution accompanied by varying degrees of evidence for CSM interaction in the spectral evolution.
The paper is organized as follows. 
In Sect.~\ref{sec:obs} we present the observations, including optical photometry and spectroscopy. The results are presented in 
Sect.~\ref{sec:results}.
Section~\ref{sec:discussion} contains a discussion and finally Sect.~\ref{sec:conclusions} presents our conclusions.

\section{Observations}
\label{sec:obs}

\subsection{First detection and classification}
\label{sec:detection}

SN\,2019zrk (also known as ZTF\,20aacbyec) was first detected on December 20, 2019 
($\mathrm{JD}_{\rm{discovery}}^{\rm{SN2019zrk}}=2458838.018$), with the Palomar Schmidt 48-inch (P48) Samuel Oschin telescope as part of the 
Zwicky Transient Facility (ZTF) survey \citep{2019PASP..131a8002B,Graham2019}. It was reported to the Transient Name Server (TNS\footnote{https://wis-tns.weizmann.ac.il/}) on  January 3, 2020. 
That first detection was in the $g$ band, with a host-subtracted magnitude of $20.18\pm0.31$~mag, at the J2000.0 coordinates $\alpha=11^{h}39^{m}47.38^{s}$, $\delta=+19\degr55\arcmin46.6\arcsec$. 

\begin{figure*}
\centering
\includegraphics[width=9cm]{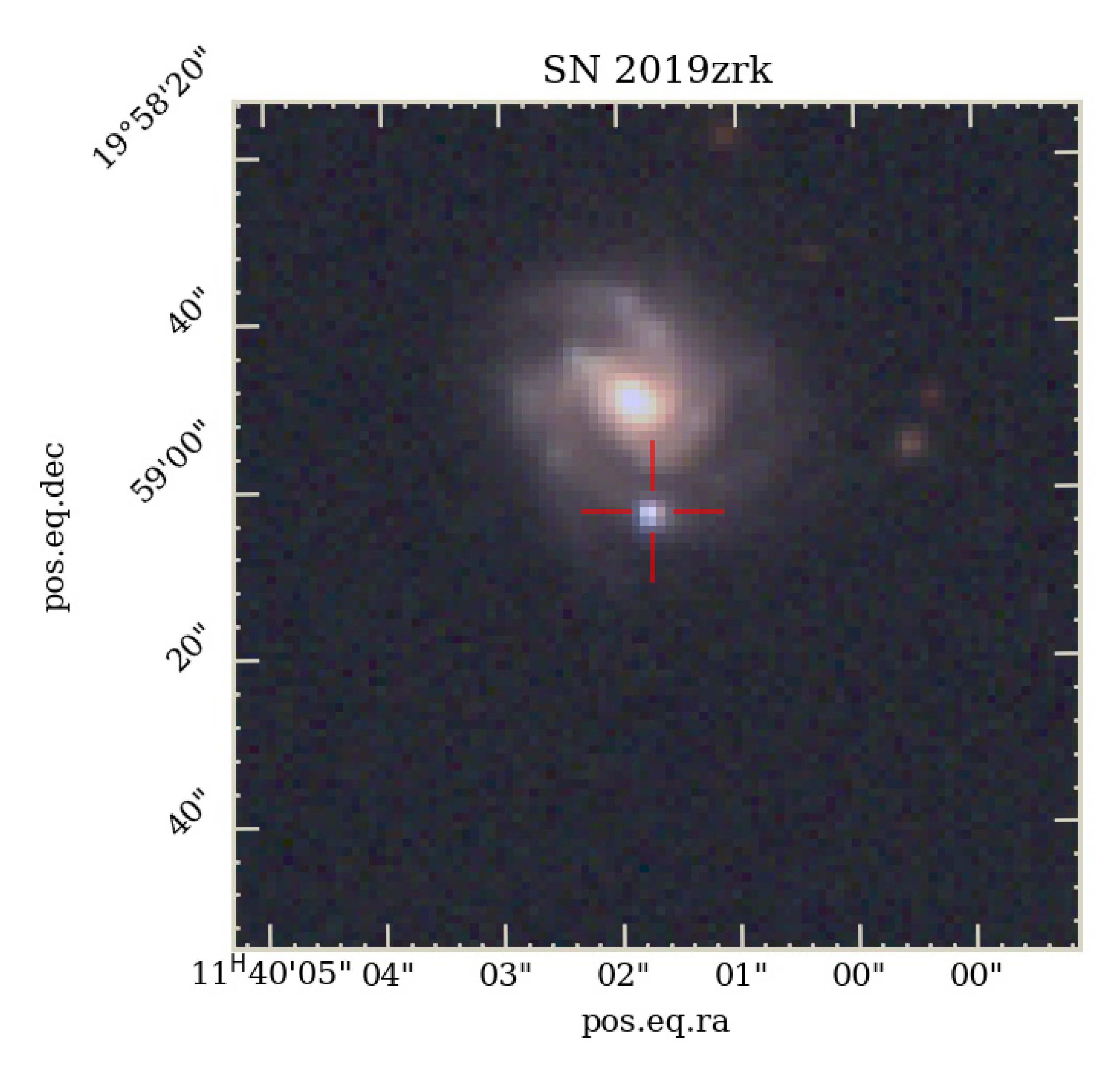}
\caption{ 
Field of SN 2019zrk and the host galaxy UGC 6625.  The image is composed of $gri$ images obtained with the ZTF camera on February 16, 2020, close to the time of peak luminosity. The supernova is marked with a red cross.
\label{fig:host}}
\end{figure*}

The intial TNS discovery report \citep{2020TNSTR..33....1D} 
reported a last nondetection of 19.3 mag during the same night, but in the $r$ band. 
The previous $g$-band 
observation was from nine nights before with a formal limit of 
19.43 mag. This transient was subsequently also 
detected and 
reported to the TNS by several other surveys: in January by Pan-STARRS, in February by ATLAS, and in April by Gaia. SN 2019zrk occurred in the spiral galaxy UGC 6625 (Fig. \ref{fig:host}) that has a reported redshift of
$z = 0.03647$ \citep{Cortese2008}. Using a flat cosmology with H$_0=70$~km~s$^{-1}$~Mpc$^{-1}$ and $\Omega_{\rm{m}} = 0.3$, this corresponds to a distance of 168 Mpc when accounting for the NED\footnote{https://ned.ipac.caltech.edu/} infall model.

\begin{figure*}
\centering
\includegraphics[width=0.8\textwidth]{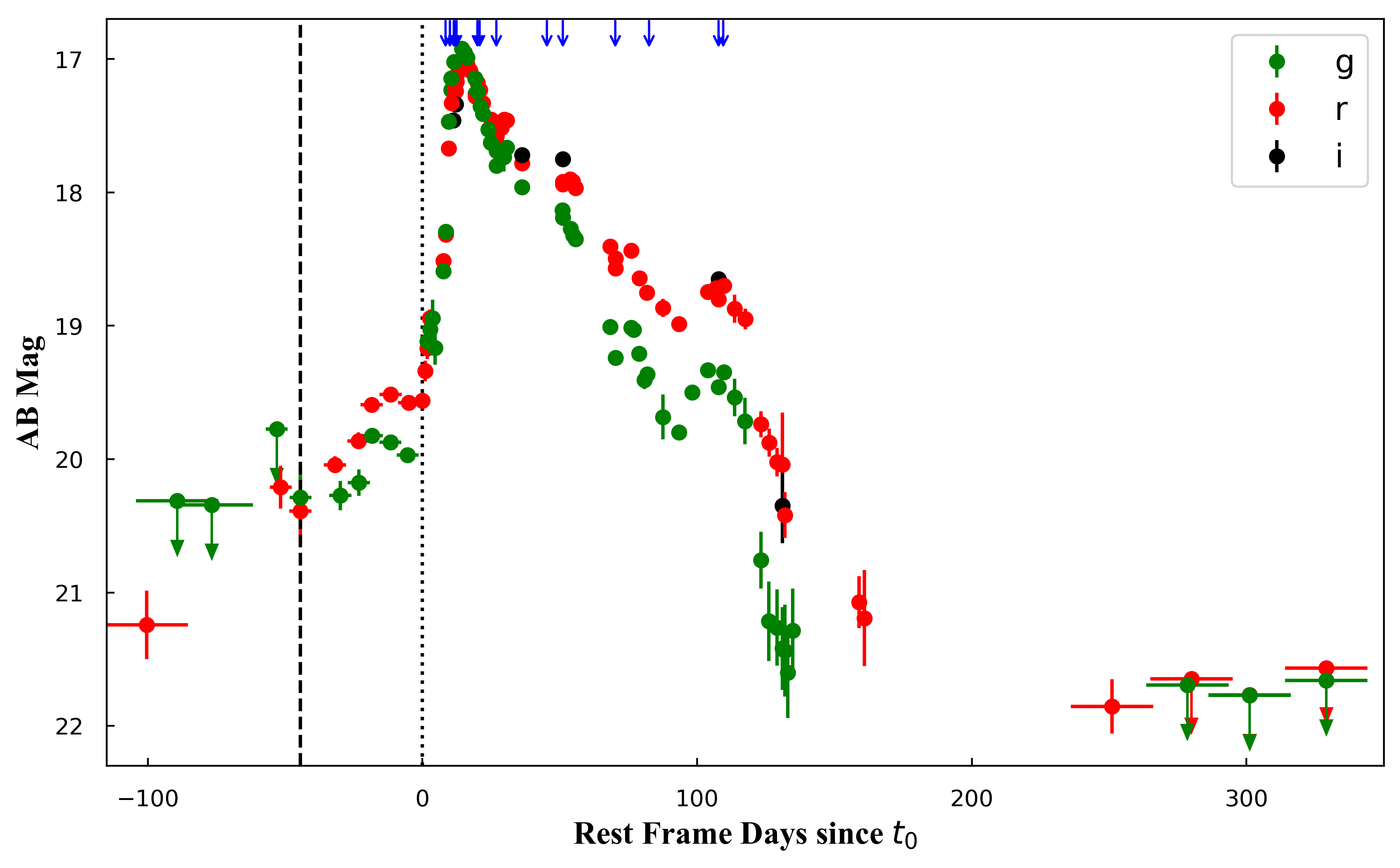}
\caption{Light curves of SN 2019zrk in the $g$ (green symbols), $r$ (red) and $i$ (black) bands. These are observed (AB) magnitudes plotted versus rest frame time in days since the rise of the main peak on
$\mathrm{JD}_{\rm{rise}}^{\rm{SN2019zrk}}=2458884$ (Sect. \ref{sec:lc1}), marked as the vertical dotted line.  The vertical dashed line marks the epoch of the first $5 \sigma$ detection of the precursor.  The horizontal error bars indicate the 7 and 30 day binning periods used for the precursor and late observations. Magnitudes are corrected for Milky Way extinction. The blue arrows on top show epochs of spectroscopy. 
\label{fig:lc}}
\end{figure*}

We classified SN\,2019zrk as a Type IIn supernova based on a spectrum obtained on 2020 February 17 with the
Dual Imaging Spectrograph (DIS) on the Apache Point Observatory (APO) 3.5-meter telescope
\citep{2020TNSCR.535....1G}. The spectrum was immediately made available on TNS. This was actually not the first spectrum we obtained, as can be seen in the spectral log (Table~\ref{tab:spec}). 
The first one was 
obtained on 2020 February 13 with
the Palomar 60-inch telescope (P60; \citealp{2006PASP..118.1396C}) equipped with the Spectral Energy Distribution Machine (SEDM; \citealp{2018PASP..130c5003B}). However, that early spectrum was not convincing enough for a proper classification.

\subsection{Optical photometry}
\label{sec:optical}

As part of the ongoing ZTF survey, we obtained regular follow-up photometry over the next 6 months in the $g$, $r$, and $i$ 
bands with the ZTF camera 
\citep{dekany2020} on the P48.
In addition, we got three epochs of $ugriz$ photometry of the object using the 
Liverpool telescope \citep[LT;][]{Steele2004},
as well as five epochs with the SEDM rainbow camera on the P60.
Forced-photometry light curves from the P48 were generated by running the pipeline described by \citet{Yao2019}, which is similar to the ZTF forced photometry service \citep{2019PASP..131a8003M}. The image subtraction is based on the \cite{Zackay2016} algorithm.
Photometry from the P60 was produced with the image-subtraction pipeline described in \cite{Fremling2016}, with template images from the Sloan Digital Sky Survey (SDSS; \citealp{2014ApJS..211...17A}). This pipeline produces PSF magnitudes, calibrated against SDSS stars in the field. We give the resulting magnitudes in Table \ref{tab:phot}.
All magnitudes are reported in the AB system.  For the early epochs of the precursor, $\la -100$ days, as well as for the very late epochs, $\ga 200$ days we have employed a 30 day binning of the observations to increase the 
signal-to-noise ratio (S/N), and for  $-100$ to 0 days we used a 7 day binning.

The light curves are shown in Fig.~\ref{fig:lc}.
We take the rise of the main peak in the $r$ band on
($\mathrm{JD}_{\rm{rise}}^{\rm{SN2019zrk}}=2458884$) as the zero point, $t_0$, for the light curve, and refer to all epochs in the rest frame of the SN with regards to this phase.

In our analysis we have corrected all photometry for Galactic extinction, using the Milky Way (MW) color excess 
$E(B-V)_{\mathrm{MW}}=0.022$~mag toward the position of SN 2019zrk
\citep{2011ApJ...737..103S}.
All reddening corrections are applied using the \cite{1989ApJ...345..245C} extinction law with $R_V=3.1$. No further host galaxy extinction has been applied, since there is no sign of \ion{Na}{id} absorption in any of our spectra. We note, however, that this method is of limited value for low resolution spectra, as these in this study \citep{Poznanski2011}.

\subsection{Optical spectroscopy}
\label{sec:opticalspectra}
Spectroscopic follow-up was conducted with a suite of telescopes and instruments. 
We most often used the SEDM mounted on the P60, but also obtained higher quality spectra with the Nordic Optical Telescope (NOT) using the Alhambra Faint Object Spectrometer (ALFOSC), 
the Keck-I telescope using the Low Resolution Imaging Spectrograph (LRIS; \citealp{1994SPIE.2198..178O}), 
as well as with the Double Beam Spectrograph (DBSP, \cite{Oke1982}) on the P200 
and SPRAT on the LT.
The first weeks of this transient were not spectroscopically covered, we only got to it with a spectrograph once it had reached peak. The early plateau was $19.5-20$ magnitudes which is a regime that is not always well covered spectroscopically by the ZTF collaboration.
A log of the spectral observations is provided in Table~\ref{tab:spec}, which includes 14 epochs of spectroscopy. 

SEDM spectra were reduced using the pipeline described by \citet{rigault} and the spectra from La Palma were reduced using standard pipelines and procedures for each telescope and instrument.
Finally all spectra were absolute calibrated compared to the host-subtracted forced photometry in the $r$ 
band, as interpolated using the Gaussian Process method.
The spectral data and corresponding information will be made available via WISeREP\footnote{\href{https://wiserep.weizmann.ac.il/}{https://wiserep.weizmann.ac.il/}} \citep{Yaron:2012aa}.

\section{Results}\label{sec:results}

\subsection{Light curves}
\label{sec:lc1}
The $g$, $r$ and $i$-band light curves of SN 2019zrk are displayed in Fig.~\ref{fig:lc}.  We do have some complementary photometry also in other bands. These data are not plotted here for clarity, but are provided in the data-files released with the paper. 
The most remarkable part of the light curve (LC) for this Type IIn SN is the initial very slow evolution - a gentle plateau, {a precursor} - of at least 50 days, before the LC starts rising in earnest. 

The rise of the main peak starts on
$t_0$ (= $\mathrm{JD}_{\rm{rise}}^{\rm{SN2019zrk}}=2458884$), with an uncertainty of $\pm 1$ day, best seen in the $r$ band. 
The rise in both the $r$ and $g$ bands occurs in two steps, first relatively slowly up to +8 days and then more rapidly up to the peak at +14 days. 
In the $g$ band, the light curve rises by
2.9 magnitudes in 
15 days. It is therefore clear that the initial plateau is a precursor to the main eruption. 
The properties of the precursor are further discussed in Sect.~\ref{sec:precuror}.

The rise times for Type IIn SNe span a wide range, and are typically
around $20\pm6$ days, with some going somewhat slower at $50\pm11$ days \citep{Kiewe2012,Nyholm20}, and occasionally some
rise even more slowly, like SN 2015da that needed 100 days to continuously rise to $r$-band peak \citep{Tartaglia20}. 
SN 2019zrk therefore rises relatively fast (not accounting for the precursor) compared to other Type IIn SNe. 

After peak, the light curve of SN 2019zrk declines in a way not untypical of some fast declining Type IIn SNe, 
but we do note a very clear bump in the light curve after $\sim 110$ days, seen in both the $g$ and $r$ LCs. A less pronounced bump or ledge is also present at $\sim 50$ days.
Such bumps are rare
\citep{Nyholm20} but there do exist clear cases of similar undulations in Type IIn supernovae
\citep[e.g.,][]{Nyholm13z}, as well as in the radio light curves of Type IIL SNe \citep[e.g.,][]{Weiler1992,Montes2000}. In Sect.~\ref{sec:comparison} we discuss the particular cases of 09ip-like light curves, where bumps are present.
These are a clear indication of circumstellar interaction in a nonuniform CSM.

\begin{figure}
\centering
\includegraphics[width=9cm,angle=0]{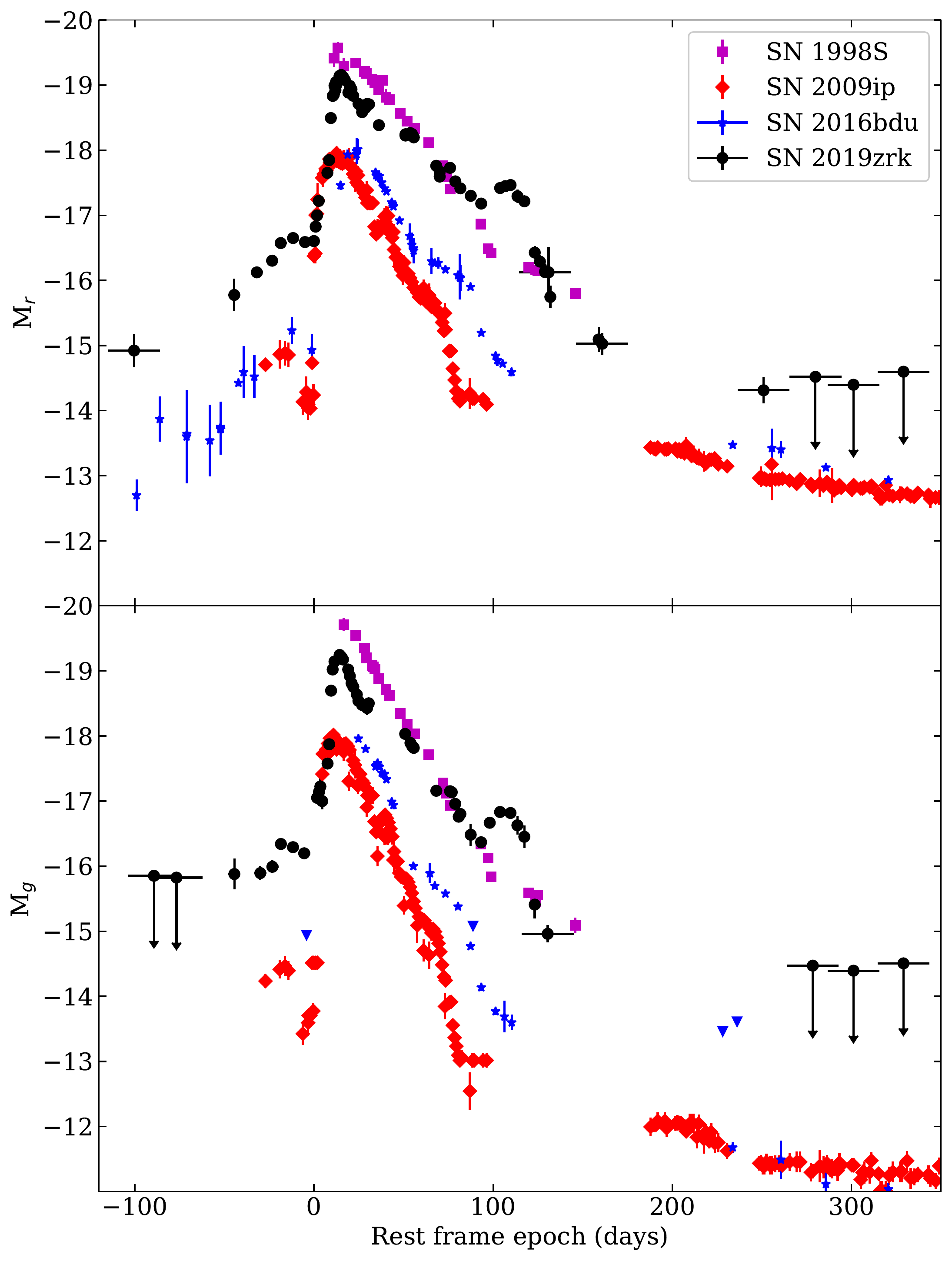}
\caption{\label{fig:comp_photom} Absolute $r$- (top) and $g$-band (bottom) magnitudes for SN 2019zrk (black dots) together with those of SNe 2009ip,  2016bdu and 1998S.
} 
\end{figure}

In Fig.~\ref{fig:comp_photom} we show the light curves in absolute magnitudes. The magnitudes 
are in the AB system, and have been corrected for distance modulus and MW extinction, $\mu=36.12$ and $E{(B-V)}=0.022$ mag, respectively, and are plotted versus rest frame days since t$_0$. 
For the discussion in Sect.~\ref{sec:discussion}, we have also included SN\,2009ip \citep{Mauerhan2013,Graham2014} 
in the figure, 
using $\mu=31.55$ and $E{(B-V)}=0.018$ mag, SN\,2016bdu with  $\mu=34.37$ and $E(B-V)=0.013$ mag \citep{Pastorello2018}, and SN\,1998S from   with  $\mu=31.15$ and $E(B-V)=0.22$ mag \citep{Fassia2000}. The light curves of SN\,2009ip and SN\,2016bdu have be shifted in time to agree with the start of the main eruption. For SN\,1998S, where this was not observed, we have synchronized the times of the peaks.
The peak brightness of SN\,2019zrk is
m$_{\rm r}^{\rm{peak}}$ = 17.0 mag 
which in absolute magnitude becomes
M$_{\rm r}^{\rm{peak}}$ = $-$19.2 mags, applying the above mentioned distance and extinction.
 
To calculate the bolometric light curve of SN\,2019zrk, we used the method from \cite{Lyman2014}, based on the $r$- and $g$-band photometry, using the epochs where we have clear detections. For epochs with only the $r$ band (later than 150 days and the first detection at $-100$ days) we use the color of the nearest epoch to calculate the bolometric magnitude.  This method is mainly intended for CC SNe without dominant CSM interaction, but most of the principles behind should at least approximately apply also to Type IIn SNe. 
To check this we have also integrated the observed spectra for the epochs where we have calibrated data (see Fig.~\ref{fig:spec}).   We note, however, that there may be a substantial contribution to the total luminosity outside the optical band in both the ultraviolet (UV) and in X-rays, as was for example seen in SN\,2009ip \citep{Margutti2014}. The light curve derived here should therefore be seen as pseudo-bolometric only.

\begin{figure}
\centering
\includegraphics[width=9cm,angle=0]{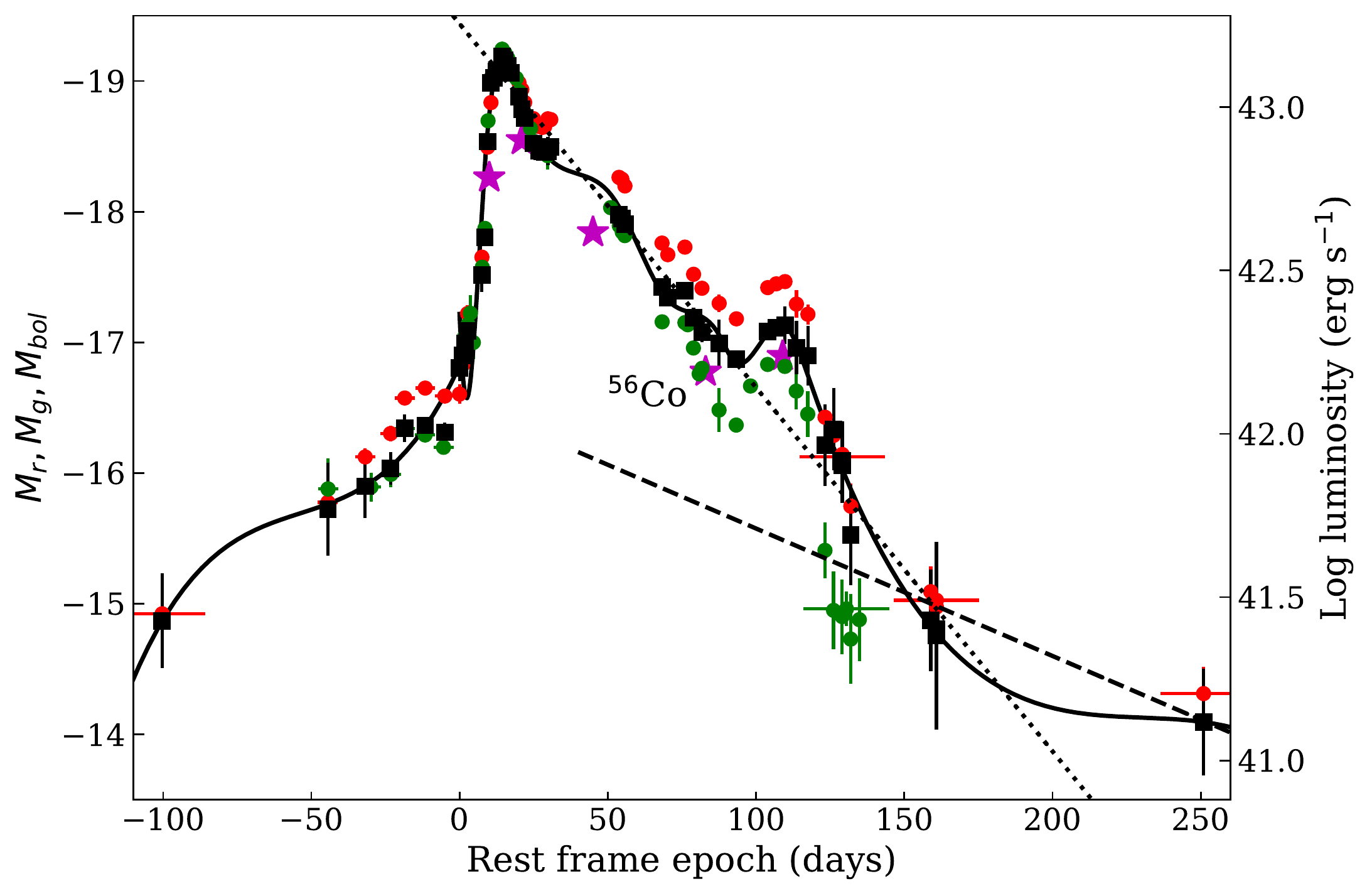}  
\caption{\label{fig:bolom} Bolometric light curve  (black squares and solid line) of 
SN\,2019zrk, together with the  $r$- (red) and $g$-band (green) photometry. The magenta stars show the pseudo-bolometric magnitudes from our spectra, while the dashed line shows the expected late light curve from ${}^{56}$Co decay, corresponding to a ${}^{56}$Ni mass of 0.09 \msun  and full trapping in the nebular stage. The dotted, black line shows an exponential fit to the luminosity (see text).
} 
\end{figure}

Figure~\ref{fig:bolom} shows the result of this exercise. 
We note the agreement between the two different methods of estimating the bolometric luminosity for the dates when spectra are available. It therefore seems as if the simple color-based method of \cite{Lyman2014} works reasonably well also for Type IIn SNe.

 \subsection{Precursor}
 \label{sec:precuror}

\cite{Ofek2014b} and \cite{nora} systematically explored the PTF and ZTF pre-discovery forced photometry searching for precursors and outbursts for Type IIn SNe. The results for SN\,2019zrk are included in Fig.~\ref{fig:lc}. 
This reveals that the precursor was present even before what we have defined as the first detection above. 
Using 7-day bins, there is a significant $r$-band detection already at day $-53$  ($\mathrm{JD}=2458829.0$), 
nine days before discovery. 
Binning the data in wider 30-day bins reveals a $3.3\sigma$ detection in the $r$ band at day $-104$ 
($\mathrm{JD}=2458780$). 
The precursor was hence potentially even $\gtrsim100$ days long (Fig.~\ref{fig:lc}).
The median absolute $r$-band magnitude of the precursor was $-16.44$, while the median $g$-band magnitude was $-16.19$. \\ 

\begin{figure}
\includegraphics[width=9cm]{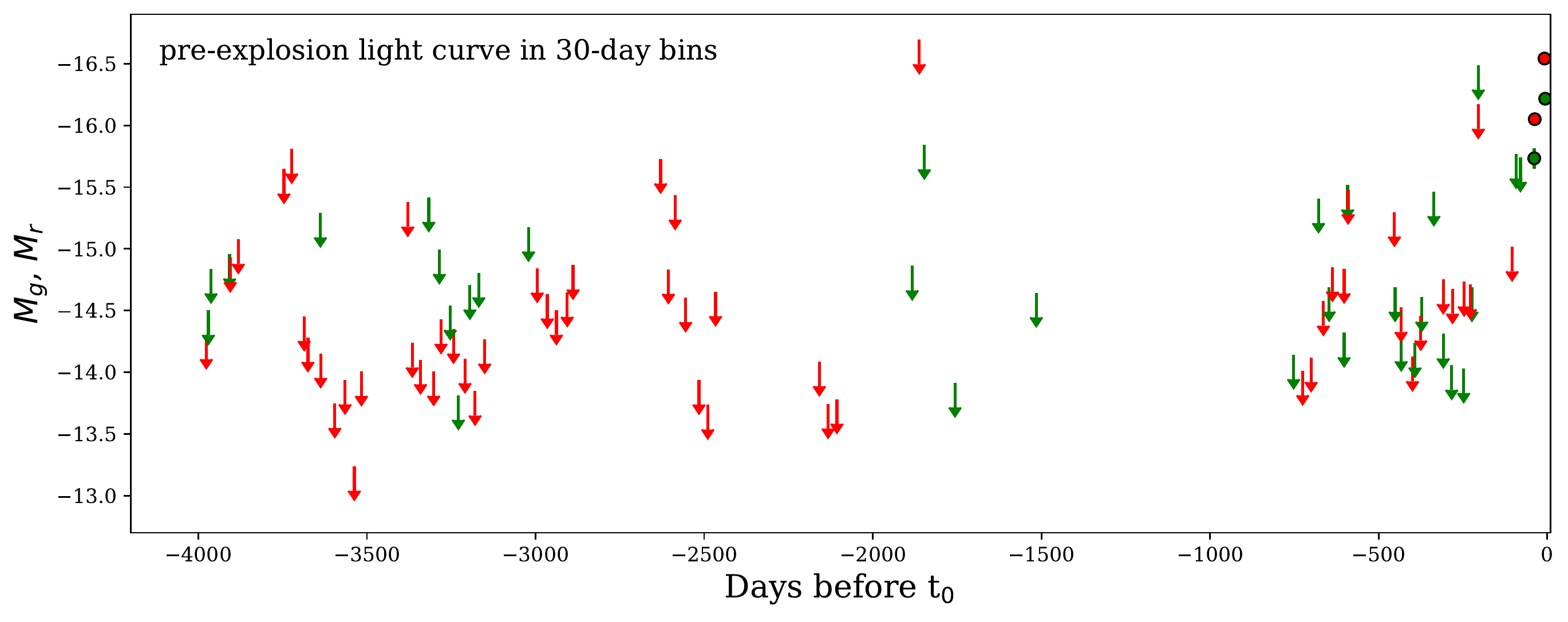}
\caption{Pre-explosion light curve. Observations up to 1\,500 days before the SN peak were obtained as part of the PTF survey, while later observations are from the ZTF survey. Except for the $\sim 100-$day long precursor before the explosion (solid data points), no earlier precursors are detected.
}
\label{fig:ptf_ztf_lc}
\end{figure}

The bumps in the fading light curve in Fig.~\ref{fig:lc} indicate interaction with different CSM shells, which points to several separate mass ejection events. We therefore searched for additional precursors at earlier times. The commissioning of the ZTF camera started in fall 2017 and we hence have data obtained over two years before the explosion. Using the methods described in
\cite{nora}, we combine observations in 1-day to 3-month long bins and search the binned light curve for detections that are significant at the 5$\sigma$ level. No additional precursors are found and Fig.~\ref{fig:ptf_ztf_lc} shows the resulting upper limits for month-long bins.

In addition, data from the Palomar Transient Factory survey allows us to search for precursors that occurred 11 to 4 years before the explosion of SN\,2019zrk. We use the IPAC PTF forced photometry pipeline~\citep{Masci2017} to produce a light curve consisting of 659 observations obtained in the SDSS $g$ and Mould-$R$ band. We remove data points that are not photometrically calibrated and do a baseline correction by subtracting the median flux from the complete light curve. Error bars are scaled up by 34\% and 14\% in the $g$ and $R$ band, respectively, such that they account for the standard deviation of the light curve. We do not find any data points that reach a significance of $\ga 5\sigma$ when searching unbinned light curves as well as light curves binned in 1-day, 3-day, 7-day, 15-day, 30-day, and 90-day bins. The complete pre-explosion light curve is shown in Fig.~\ref{fig:ptf_ztf_lc}.

Assuming that precursors last for at least one month we can quantify the amount of time during which our observations, combined in month-long bins, exclude the presence of a precursor. With a median absolute magnitude of $-16.4$ in the $r$ band the detected precursor just before the main peak was quite bright (see Fig.~\ref{fig:ptf_ztf_lc}) and such a bright precursor would have been detectable during 55 months in the 11 years before the SN main eruption or during 40\% of the time. Fainter precursors with an absolute magnitude of $-15$ can still be ruled out 32\% of the time, while precursors with an absolute magnitude of $-14.5$ would have been detected only 19\% of the time. Even fainter events are typically below the sensitivity threshold of our search and would remain undetected.

We observe that the precursor first becomes redder until day $-4$,
which could either be due to a cooling continuum or due to the strengthening of the $\text{H}\alpha$ line which falls in the $r$ band. Then, as the light curve rises (see Fig.~\ref{fig:lc}), the $g-r$ color index becomes bluer again, approximately until the SN reaches its peak. We hence observe that the increase in flux is associated with bluer colors.

\subsection{Spectroscopy}\label{sec:spec}

\begin{figure*}
\centering
\includegraphics[width=17cm,angle=0]{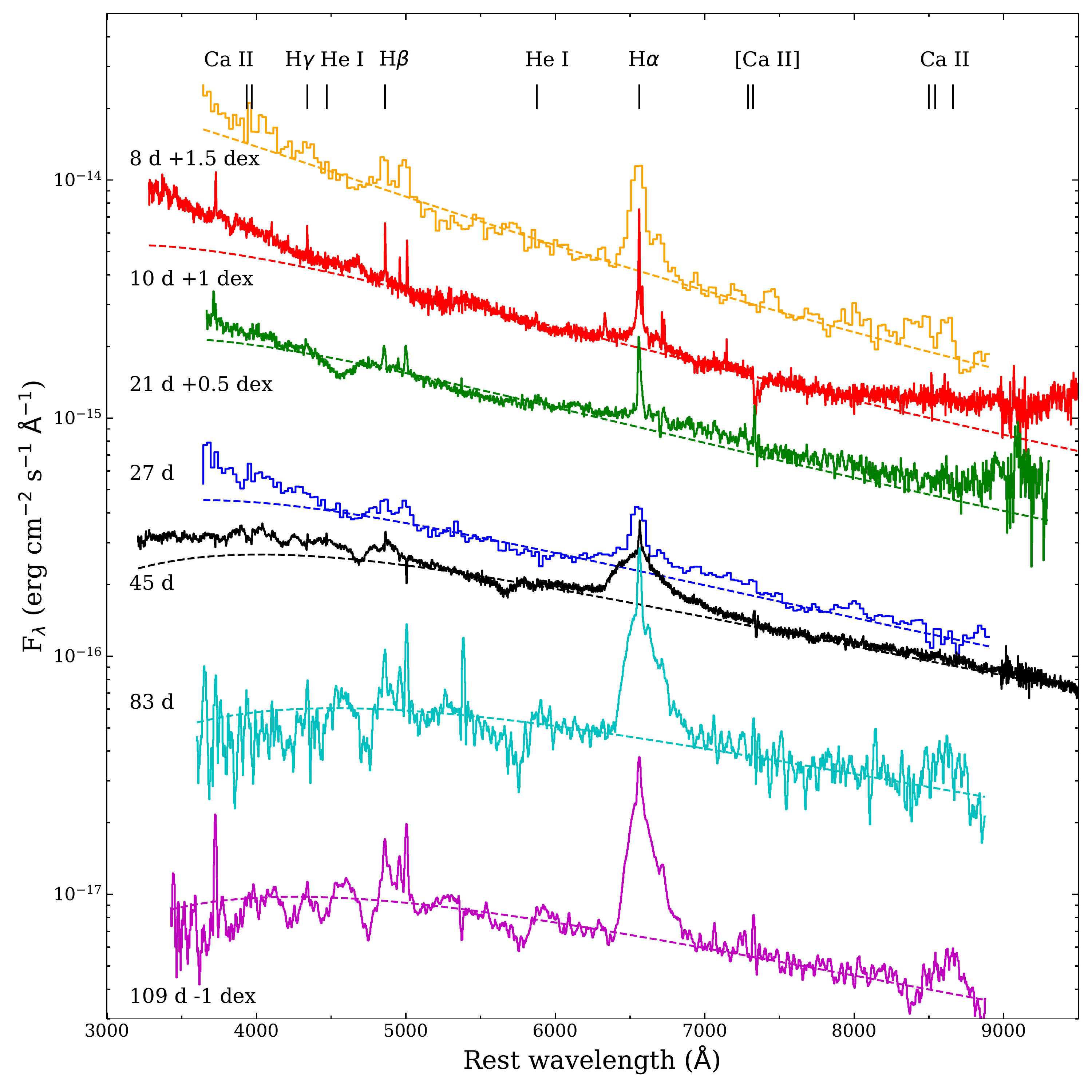}  
\caption{\label{fig:spec} Spectral sequence illustrating the spectral evolution of SN\,2019zrk. The early P200 spectrum at 8 days (red) revealed a Type IIn SN with a blue continuum and superimposed narrow emission lines.
A month later, the deep Keck spectrum (black) instead shows a broad and asymmetric line profile of H$\alpha$, which typically reflects strong CSM interaction. Finally, the day 83 and 109  NOT spectra display increasingly strong P-Cygni absorption features. The dashed lines for each spectrum show blackbody fits to the continuum (see text).
For clarity some of the spectra have been shifted by the logarithmic factor provided together with the phase in the figure. All spectra are corrected for the Milky Way extinction.
}
\end{figure*}

The classification spectrum of SN\,2019zrk revealed a Type IIn SN (Sect.~\ref{sec:detection}). 
As mentioned, we had a sequence of spectra already before the classification spectrum, although the first one was obtained when the SN was already close to peak brightness. The epochs of spectroscopic observations are illustrated in the light-curve figure (Fig.~\ref{fig:lc}) and the log of spectroscopic observations is provided in Table~\ref{tab:spec}.

As can be seen in the spectral sequence (Fig.~\ref{fig:spec}), SN\,2019zrk underwent an interesting 
spectral evolution. Early spectra revealed a Type IIn supernova with a blue continuum and superimposed narrow emission lines with faint broad wings, characteristic of electron scattering and therefore strong CSM interaction. The line core can in low resolution spectra be difficult to distinguish from the narrow emission lines from the host galaxy. 
Later on, at 45 days a deep Keck spectrum showed a broad and asymmetric line profile of H$\alpha$.
The last spectra at 83 and 109 days still have a strong H$\alpha$, but now also with strong absorptions in H$\beta$, He I, and Ca II.

\begin{figure}
\centering
\includegraphics[width=10cm,angle=0]{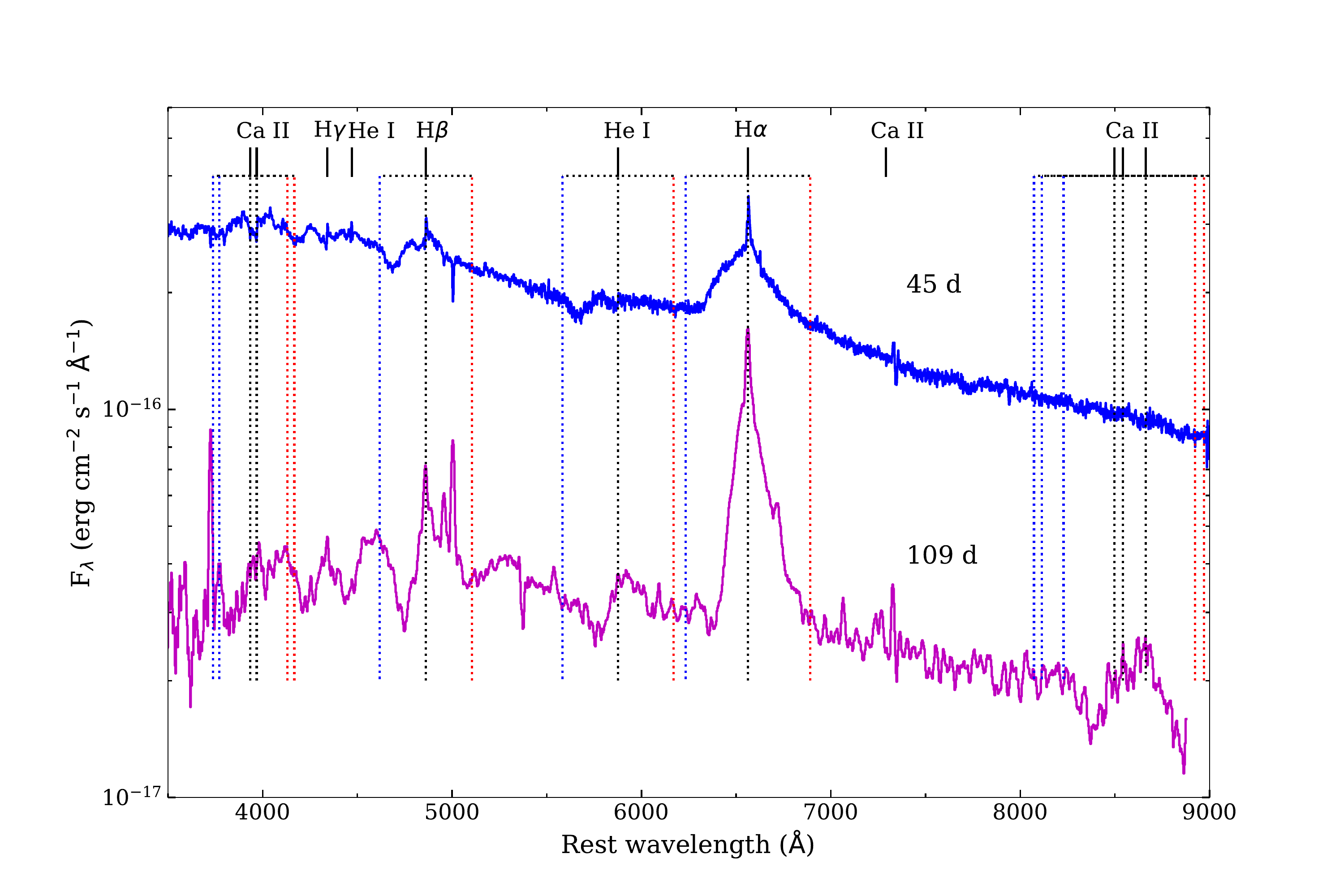}  
\caption{\label{fig:spec_90_154} Comparison of the spectra at 45 (blue) and 109 (magenta) days. We note the increase in the relative H$\alpha$ flux and the much deeper absorption features in the day 109 spectrum. The dotted blue and red vertical lines show the wavelengths corresponding to $\pm 15,000 \kms$. The spectra have been smoothed by a third order Savitzky-Golay filter, with windows of 5 and 11 pixels, respectively. } 
\end{figure}

To illustrate the late evolution, Fig.~\ref{fig:spec_90_154} shows the 45 and 109 days spectra, which are the spectra with best S/N
at late epochs. The 45 day spectrum shows most characteristics of CSM interaction with a hot spectrum dominated by the continuum and an H$\alpha$ line dominated by scattering.  While most of the line features are present also in the 109 day spectrum, the weak P-Cygni absorptions present in the day 45 spectrum have become very strong and prominent. The H$\alpha$ line has also increased strongly relative to the continuum and has nearly the same flux as at 45 days. 

The P-Cygni absorptions in the late spectra can be used to estimate the maximum velocity of the expanding ejecta. In Fig.~\ref{fig:spec_90_154} the blue and red dotted vertical lines show the wavelengths corresponding to $\pm 15,000 \kms$, respectively.  While for \Ha \ the P-Cygni absorption is almost filled-in by the emission, the absorption is very clear in the 83 and 109 day spectra for H$\beta$, as well as for H$\gamma$, He I $\wll 4471, 5876$, the permitted Ca II H and K lines, and the Ca II NIR triplets.
The exact extension of the absorptions, and therefore the maximum velocity of the ejecta, is somewhat difficult to determine. The blue velocity marked at $15,000 \kms$ is a minimum velocity. 

\begin{figure*}
\centering
\includegraphics[width=15cm,angle=0]{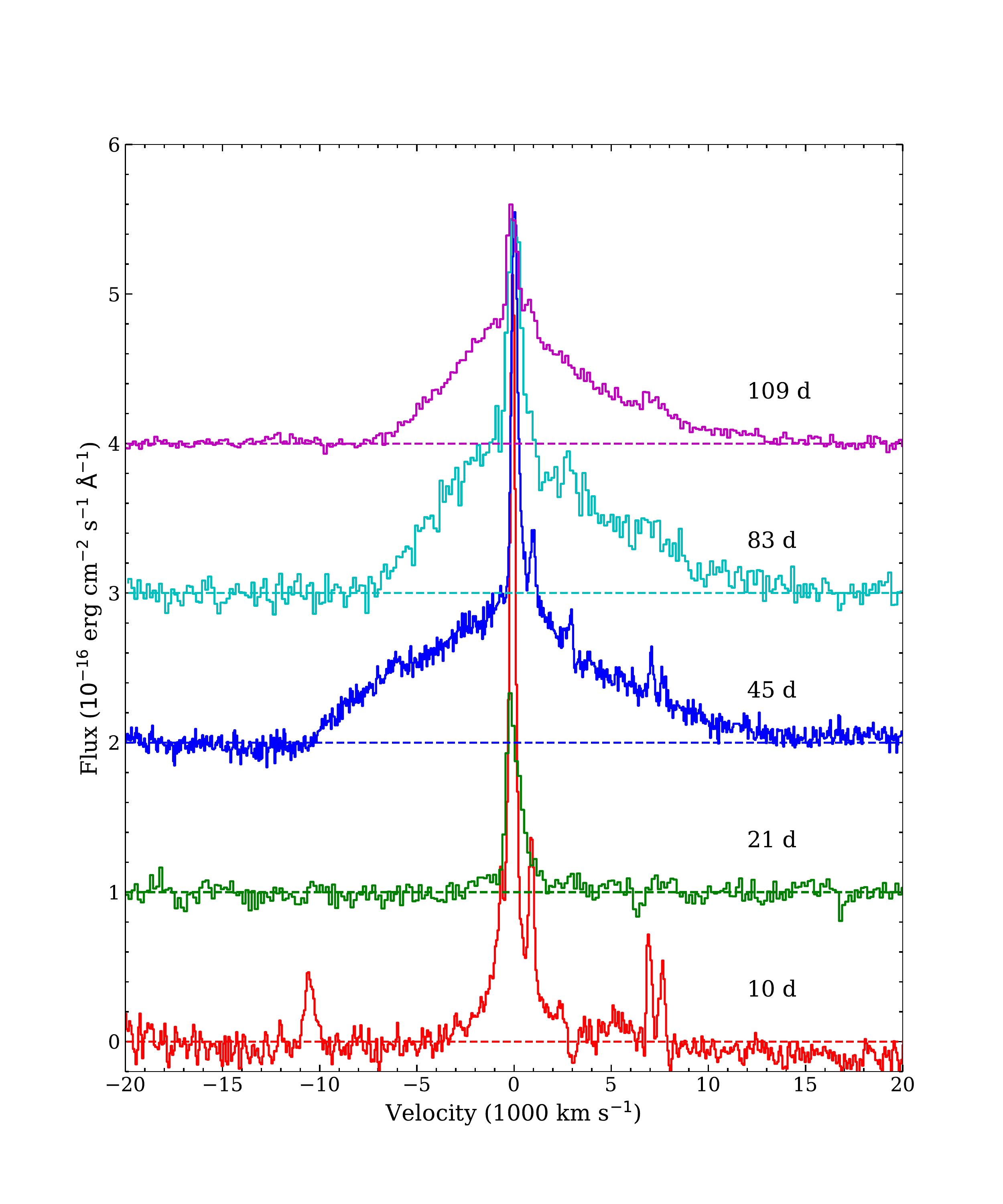}  
\caption{\label{fig:Halpha_evol} Evolution of the H$\alpha$ line. The continuum level is subtracted and each spectrum is shifted upwards for clarity. Note the dramatic change in line profile between days 21 and 45, where the narrow line is replaced by a broad line profile, reflecting the shock velocity. Each spectrum has been shifted by $10^{-16}$ erg cm$^{-2}$ s$^{-1}$.  
}
\end{figure*}

In Fig.~\ref{fig:Halpha_evol} we show the evolution of the H$\alpha$ line in detail. We have subtracted the background continuum by fitting a second-order polynomial to the line-free parts of the spectrum between $-20,000 - -15,000 \kms$ \ on the blue side and between $+20,000 - +50,000 \kms$ \ on the red side. We here see a dramatic change in the line profile between days 21 and 45, where the narrow line is replaced by a broad line profile. Before the transition the line is dominated by a narrow core with FWHM $\sim 220 \kms$ at 10 days, consistent with the resolution of the P200/DBSP with the 600 lines/mm \  grating. There are, however, in both the 10 and 21 day spectra faint wings seen to $\sim 5000 \kms$, although the maximum velocity shift seen is set by the S/N. We note that this is a result of electron scattering and not of the expansion velocity of the ejecta.

Between the day 21 and day 45 spectra the wings become very strong, and also asymmetric. This can also be seen in the 
low resolution P60 spectrum from day 27. In the higher resolution Keck spectrum with the best S/N, the line profile shows a blue wing to $-10,500 \kms$, while the red wing extends to at least $+13,500 \kms$, determined mainly by the exact level of the continuum. The blue wing shows a "kink" at $\sim 6000 \kms$, where the profile becomes somewhat steeper.  In the last two spectra at 83 and 109 days the maximum blue-shift decreases to $\sim 7500 \kms$, while the red wing extends to a fairly constant velocity of $13,000-15,000 \kms$. 

\begin{figure}
\centering
\includegraphics[width=9cm,angle=0]{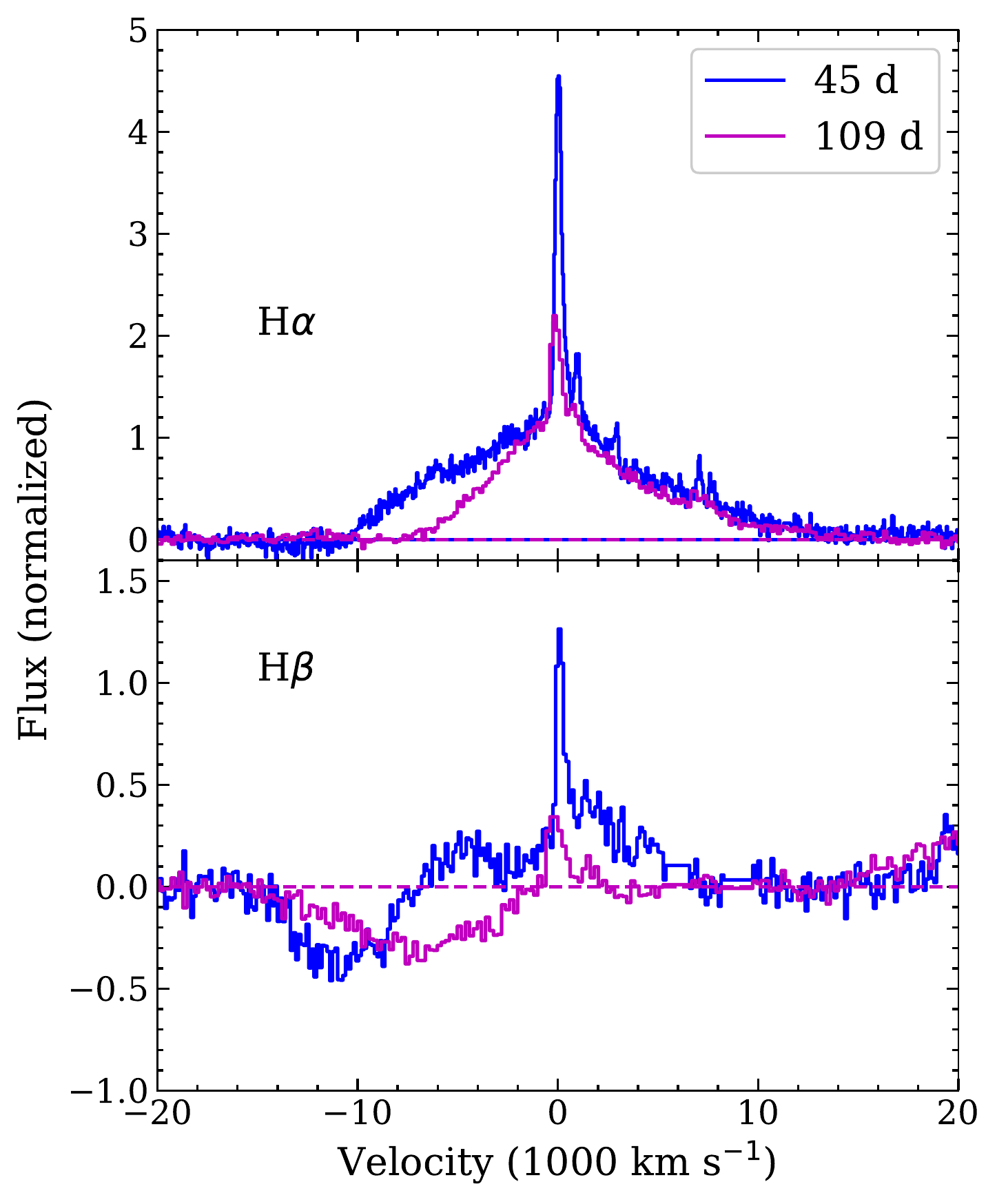}  
\caption{\label{fig:Halpha_compl} Comparison of the H$\alpha$ (upper panel) and H$\beta$ (lower) line profiles at 45 (blue) and 109 (red) days. The flux of both lines and at both epochs have been normalized to the \Ha \ flux density at $-2000 \kms$.  We note the nearly identical red wing of \Ha, while the blue wing above $\sim 4000 \kms$ shows a strong decrease in flux.  The \Hb \ line shows a similar evolution of the blue wing, with an increasing absorption component of the P-Cygni profile. The [O III] $\lambda\lambda 4959, 5007$ lines from the background in the red wing of \Hb \ have been masked out.
}
\end{figure}

In Fig.~\ref{fig:Halpha_compl} we compare the \Ha \ and H$\beta$ line profiles at 45 and 109 days. Here we have normalized these to the \Ha \ flux at $-2000 \kms$. The comparison shows that while the red wing of \Ha \ is nearly constant, the blue wing displays a strong decrease in the flux above $\sim 4000 \kms$ between 45 and 109 days, which can be seen already at 83 days (Fig.~\ref{fig:Halpha_evol}). However, after that the line profile is nearly constant.

The H$\beta$ profile is especially interesting because it shows a strong P-Cygni absorption at velocities $\ga 6000 \kms$ at 45 days, with a minimum at $\sim 10,000 \kms$. At this epoch it also has a weak red emission component. At 109 days the emission component has almost completely disappeared, while the blue absorption now extends from $\sim -16,000 \kms$ \ to zero velocity. The increased absorption in \Hb \ is directly correlated in velocity with the decrease in the \Ha \ profile, showing that absorption in combination with decreased scattering in this line is responsible for the shape.
Figure~\ref{fig:Halpha_compl} also illustrates the increasing \Ha/\Hb \ ratio between these epochs. For the 109 day spectrum the total \Hb \  flux is actually negative, as should be the case for a P-Cygni profile dominated by scattering.  

\subsection{The host galaxy}
\label{sec:host}
Figure \ref{fig:host} shows the position of the SN and the host galaxy, UGC 6625, obtained with the ZTF camera. The database of the Sloan Digital Sky Survey contains spectra of two different regions of the host galaxy. After correcting the science-ready spectra for Milky Way extinction, we fit the H$\alpha$, H$\beta$, [O \textsc{iii}]$ \lambda$5007 and [N \textsc{ii}] $\lambda$6585 emission lines by Gaussian profiles to measure the line fluxes (Table \ref{tab:host_eml}). Using the O3N2 metallicity indicator and the parameterization by \citet{Marino2013a}, 
we infer  high oxygen abundances between 8.45 and 8.49.

To put these values in context, we need the galaxy's total stellar mass. We retrieved the science-ready images from \textit{GALEX}, SDSS, Pan-STARRS (PS1), 2MASS, and WISE surveys and measured the brightness from the FUV to NIR following \citet{Schulze2021a} (Table \ref{tab:host_phot}). Afterwards, we modeled the spectral energy distribution (SED) with the software package \texttt{prospector} version 0.3 \citep{Johnson2021a} using the same model for the galaxy SED as in \citet{Schulze2021a}. The SED is adequately described by a galaxy template with a mass of $\log\,M/M_\odot=10.84^{+0.11}_{-0.38}$ and a star-formation rate of $12.7^{+8.8}_{-2.3}~M_\odot\,\rm yr^{-1}$.
Using the mass-metallicity relation by \citet{Sanchez2017a} for the O3N2 metallicity scale and the \citet{Marino2013a} parameterization, we conclude that the oxygen abundance is high and in the expected range for such an evolved galaxy.

\section{Discussion}\label{sec:discussion}

\subsection{Light curve}
\label{sec:lc}
	
To obtain a limit of the ejected ${}^{56}$Ni mass, we show in Fig.~\ref{fig:bolom} the expected energy input due to $^{56}$Co decay as a dashed line. The normalization, set by the observations later than $\sim 100$ days, corresponds to a ${}^{56}$Ni  mass of $\sim 0.09$ \msun.  We note that the slope  before 160 days is considerably steeper than that expected for the  ${}^{56}$Co decay. For the epochs later than 150 days we can not exclude a contribution from  ${}^{56}$Ni, but a flattening may also be explained as a result of circumstellar interaction. 

In principle, the explosion could start with the ‘precursor’ and the CSM interaction starting with the main peak.  This, however, seems like a rather contrived scenario with a fair amount of fine tuning. or example, it would be difficult to understand the bumps and the dip in the light curve just before the main peak, seen in several of these 09ip like objects, as discussed in Sect. \ref{sec:Origin}.

We therefore conclude that radioactive input is likely of minor importance for the observed light curve. Except for the bump, we can instead fit the bolometric light curve very well with a steeper exponential, 	\begin{equation}
    L(t) = 1.8\times 10^{43} e^{-t/39 {\rm \ d}} \ {\rm erg \ s^{-1}}, 
    \label{eq:bolfit}
\end{equation}
shown as a dotted line in Fig. \ref{fig:bolom}. We have also tested a power-law fit, $L(t) \propto t^{-\alpha}$, which fits the light curves of long lasting Type IIn SNe \citep[e.g.,][]{fransson10jl}. This, however, represents a considerably worse fit.

The total radiated  energy from the bolometric light curve during the main eruption is $5.0 \times 10^{49}$ erg, compared to $3.7 \times 10^{48}$ erg during the precursor.  However, as demonstrated for SN\,2009ip  \citep{Margutti2014}, there may also be a substantial additional contribution from UV and X-rays.   
	
\subsection{Spectrum}
\label{sec:spec_disc}
	
In spite of the low resolution, the first spectrum at eight days with the SEDM on the P60 shows a clear and strong \Ha \ line. The first higher resolution spectrum with the P200 shows this as a narrow \Ha \ with FWHM $\sim 220\kms$, consistent with being unresolved. 
In the Keck spectrum at 45 days the width of the narrow component has increased to $\sim 690 \kms$ on top of a broader component. This broad component is a clear indication of emission from a dense CSM, while the emission from the shock and ejecta are smoothed into a 'continuum' by the optically thick electron scattering. 
As the column density of the ionized CSM decreases, and therefore the electron scattering optical depth gets lower, the \Ha \ line profile becomes clearly broadened and skewed to the blue, as is seen in the day 45 spectrum (Fig.~\ref{fig:Halpha_evol}). The emission is now reflecting the velocity of the shock. An illustration of this evolution from symmetric to blueshifted emission line profiles was also seen for the Type IIn SN 2013L \citep{Andrews2017,Taddia2020}, and was in Taddia et al. modeled with  Monte Carlo simulations of the electron scattering in an expanding shell. 

The \Ha \ line profile  on day 45 extends to at least $15,000 \kms$ on the red side. Exactly how far is sensitive to the assumed continuum level. This is larger than expected because of occultation by the photosphere, but is a natural effect of electron scattering in the ejecta. Figure~\ref{fig:Halpha_evol} shows that the extent on the red side is nearly constant up to the last spectrum at 109 days, while the extent of the emission on the blue side is decreasing from $\sim 10,000 \kms$ to $\sim 6000 \kms$ during the same time. This is a result of the decreasing ratio of emission to absorption on the blue side. 

\begin{figure}
\centering
\includegraphics[width=9cm,angle=0]{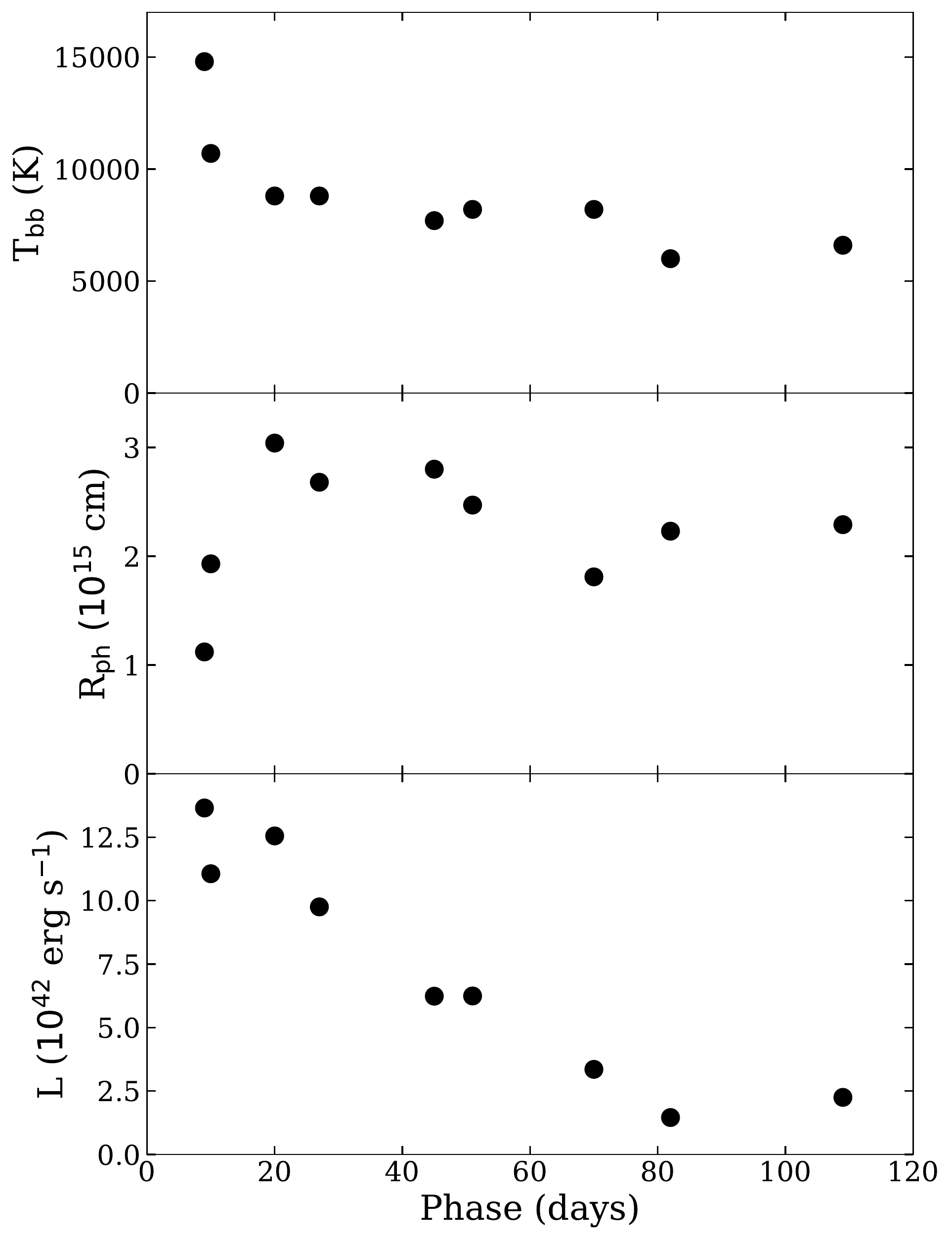}  
\caption{\label{fig:teff_rph_luml} Temperature, photospheric radius and luminosity as function of the rest frame phase from blackbody fits to the spectra. 
}
\end{figure}

Although a questionable assumption for a scattering dominated atmosphere \cite[see e.g.,][]{Dessart2015}, a blackbody fit to the continuum can give some idea of the evolution of the photospheric temperature, radius, and blackbody luminosity. We have fit Planck functions to the line free regions of our spectra, corrected for extinction, and show these as dashed lines in Fig.~\ref{fig:spec}. To calculate the luminosity we have integrated the full Planck function. This  may overestimate the luminosity because of line blanketing in the UV, especially when the temperature is $\la 10,000$ K. However, objects showing strong circumstellar interaction can often have strong line emission in the whole UV range, especially at late phases as seen in for example SN 2010jl \citep{fransson10jl} and SN 2009ip \citep{Margutti2014}. Without spectra in the UV we just note that the systematic errors may be considerable.  
The results for the blackbody temperature, $T_{\rm bb}$, photospheric radius, $R_{\rm ph}$, and luminosity, $L$, are provided in Fig.~\ref{fig:teff_rph_luml}. 

The evolution of these parameters is similar to what is seen for other interacting SNe, for example SN 2009ip \citep{Margutti2014}, with a decreasing temperature from $\sim 15,000$ K to $\sim 6000$ K at 100 days and a photospheric radius first increasing during the first 20 days and then stagnates at $(2-3) \times 10^{15}$ cm. Calculating the  photospheric velocity as $V_{\rm}= R_{\rm ph}/(t-t_0)$, and with $R_{\rm ph} \sim 3\times 10^{15}$ cm at 20 days we get $V_{\rm} \approx 17,000 \kms$. Although this ignores the initial radius, it agrees well with the maxiumum velocity seen from \Hb, $\sim 16,000 \kms$ (Sect. \ref{sec:spec}) .
 The blackbody luminosity agrees well with that in Fig.~\ref{fig:bolom}, although especially at the late phases the strong \Ha \ line and uncertain UV flux make the bolometric luminosity uncertain. 

\begin{figure}
\centering
\includegraphics[width=9cm,angle=0]{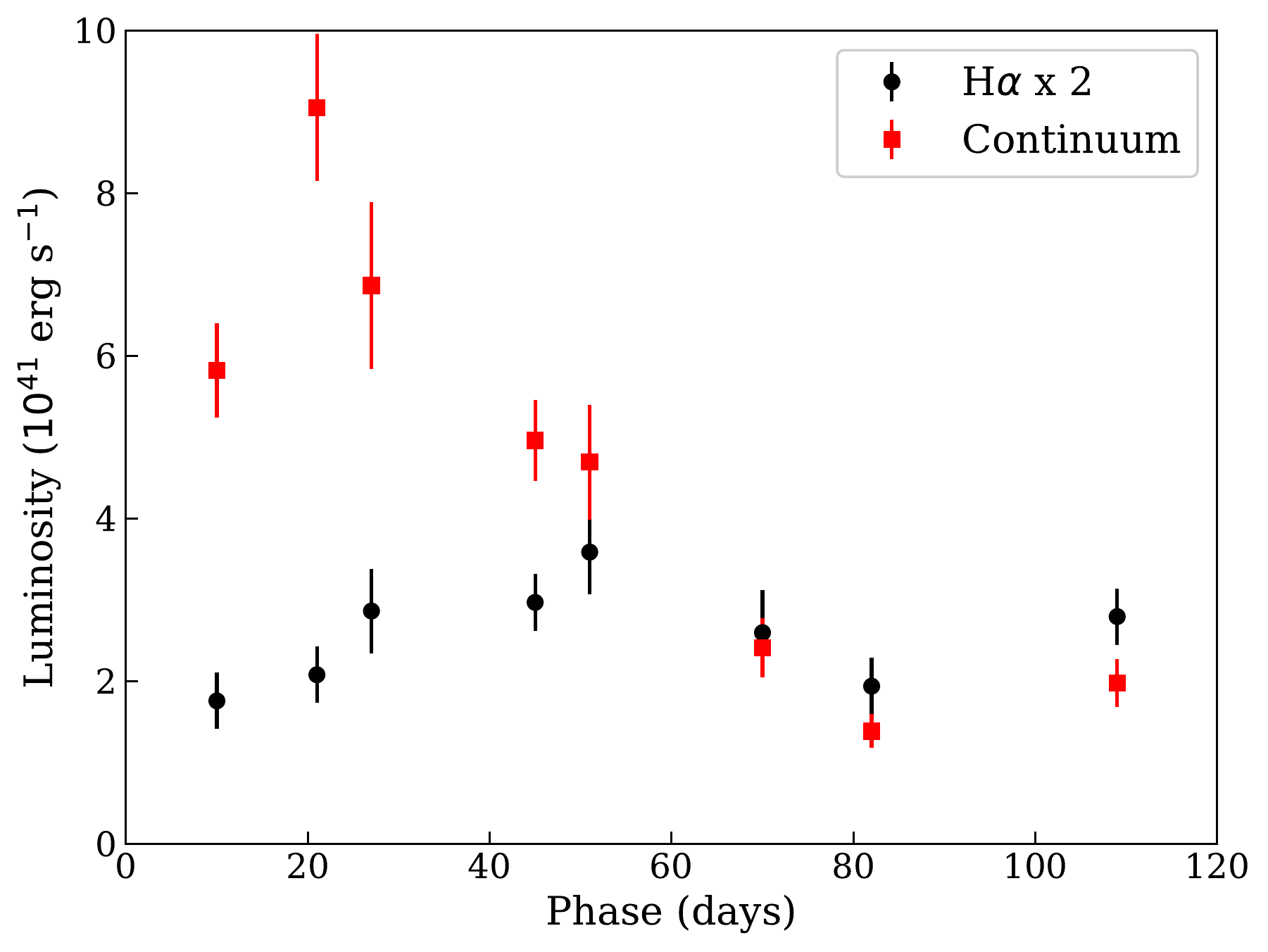}  
\caption{\label{fig:Halpha_lum} Luminosity of the H$\alpha$  line and the continuum luminosity in the range $\pm 20,000 \kms$ from the  H$\alpha$ line. For clarity the luminosity of H$\alpha$ is multiplied by a factor 2. We note the different evolution of the continuum and the H$\alpha$ luminosities.  
}
\end{figure}

In Fig.~\ref{fig:Halpha_lum} we show the evolution of the H$\alpha$ luminosity together with the continuum luminosity  in the range $\pm 20,000 \kms$ from the  H$\alpha$ line. For this we have subtracted the continuum, fit as a power law around the H$\alpha$ line. The error bars on the  H$\alpha$  luminosity depend mainly on the continuum fit and the spectral resolution, where the luminosity estimated from the SEDM spectra on days 27, 51.1, and 70 may have been slightly overestimated due to the lower spectral resolution and difficulty in subtracting the background. This is reflected in the larger error bars for these epochs.

The most interesting result 
from Fig.~\ref{fig:Halpha_lum}
is the  different evolution of the continuum and H$\alpha$ luminosities. The continuum evolution reflects the general $r$-band evolution in 
Fig.~\ref{fig:bolom}. The H$\alpha$ luminosity is initially very weak and only represents the narrow, central core of the line from unscattered H$\alpha$ photons. The luminosity then increases as the broad wings become visible at $\sim 30$ days (Fig.~\ref{fig:spec}). After $\sim 45$ days the H$\alpha$ luminosity stays nearly constant up to the last spectrum at 109 days. 

From this evolution it is clear that the H$\alpha$ emission is decoupled from the thermalization photosphere and mainly reflects the energy input from the shock. This is also indicated by the broad \Ha \ line and the decreasing importance of electron scattering in shaping the line profile. If \Ha \ is powered by the extreme-UV and X-rays from the shock, the constant level of the H$\alpha$ luminosity can be explained as a result of the fairly constant X-ray luminosity from a radiative shock, expected in an $\rho \propto r^{-2}$ medium. In this case the luminosity  $L \propto \dot M V_{\rm s}^3/u_{\rm w}$, where $\dot M$ is the mass-loss rate, $V_{\rm s}$ the shock velocity, and  $u_{\rm w}$ wind velocity. Because $V_{\rm s}$ only decreases slowly with time this results in a flat or slowly decreasing evolution for $L$.

\subsection{Estimate of the mass-loss rate}\label{sec:masslossestimate}

The observed shift in the nature of the spectrum at $\sim 50$ days (Sect.~\ref{sec:spec}) indicates a transition from a CSM which is optically thick to an optically thin medium with respect to electron scattering, as measured from the outer boundary, $R_{\rm out}$, to the shock at $R_{\rm s}$. Below we show how this can be used to obtain an estimate of the column density and the mass-loss rate. 

Assuming a steady wind the electron density is 
\begin{equation}
n_{\rm e} = \frac{\dot M x_{\rm e}}{4 \pi \mu r^2 u_{\rm w} m_{\rm p}} \ ,
\end{equation}
where $\dot M$ is the mass-loss rate, $\mu$ is the mean atomic weight, and $x_{\rm e}$ the electron fraction $n_{\rm e}/n_{\rm ion}$. For a gas with solar composition $\mu \approx 1.3$, whereas for a He-dominated gas $\mu \approx 4$. The early spectra indicate $x_{\rm e} \approx 1$, but it may be higher in the optically thick, ionized region. We scale the wind velocity to $u_w=100 \kms$, noting that the spectral resolution of the best spectrum only provides an upper limit to the unshocked, outer CSM velocity of $u_{\rm w} \la 220 \kms$.

At the time of the transition at $t_{\rm trans}\approx 50$ days the CSM in front of the ejecta should have $\tau_{\rm e} \approx 1$, and the \ion{H}{I}, 
\ion{He}{I}, \ion{Na}{I}, 
absorption lines should be formed outside this. We note that this does not correspond to the shock radius at the time of shock break-out, which occurs much earlier, when $\tau_{\rm e} \approx c/V_{\rm s} \sim 20$. 
With an ejecta velocity of $1.6 \times 10^4 \kms$ the transparency radius corresponds to $R_{\rm s}(t_{\rm trans}) \sim 7 \times 10^{15} (t_{\rm trans} / 50 {\rm \ days})$ cm. We note that this radius corresponds to an electron scattering optical depth of $\tau_{\rm e} \sim 1$ and not of the thermalization radius estimated from the blackbody radius in Fig. \ref{fig:teff_rph_luml} of $R_{\rm ph}\sim (2-3)\times 10^{15}$ cm. 

We  can now estimate the mass-loss rate
from 
\begin{equation}
\tau_{\rm e} = \frac{\dot M x_{\rm e} \sigma_{\rm T}}{4 \pi \mu u_{\rm w} m_{\rm p} } ~[\frac{1}{R_{\rm s}(t_{\rm trans})} - \frac{1}{R_{\rm out}}] \approx 1 \ ,
\end{equation}
where $\sigma_{\rm T}$ is the Thomson cross section, and  $m_{\rm p}$ the proton mass. Assuming $R_{\rm s}(t_{\rm trans}) \ll R_{\rm out}$ we get 
\begin{eqnarray}
\dot M &=& 4.4 \times 10^{-2} \left(\frac{\mu}{1.3}\right) x_{\rm e}^{-1} {\left(\frac{u_{\rm w}}{100 \  {\rm km \ s^{-1}}}\right)} {\left(\frac{t_{\rm trans}}{50  
 {\rm \ days}}\right)} \nonumber \\
&&{\left(\frac{V_{\rm s}}{16,000 \  {\rm km \ s^{-1}}}\right)}  \  \rm M_{\odot} \ year^{-1} \ .
\end{eqnarray}
Because the assumption of $\tau_{\rm e}=1$ when the CSM in front of the shock becomes transparent is only approximate,  this estimate will also only give an approximate estimate of the mass-loss rate, with an uncertainty of at least a factor two. It is, however, independent of other assumptions and uncertainties, like  bolometric corrections and the complexity going into  light-curve modeling \citep[e.g.,][]{Sorokina2016}.  

This $\dot M$ corresponds to an electron density 
\begin{equation}
n_{\rm e} = 7.5 \times 10^{8} \left(\frac{r}{10^{15} \ \rm cm}\right) ^{-2} \ \rm cm^{-3} \ ,
\end{equation}
and should be the characteristic electron density close to the photosphere at the time when the ejecta emerge through the photosphere. The gas we see, producing the narrow line cores, should be located at larger radii at a lower density.  

The  mass in front of the shock at 50 days is 
\begin{eqnarray}
M_{\rm CSM} &=& \frac{\dot M R_{\rm s}(50 \ {\rm days})}{u_{\rm w}}  \approx \\
&&0.96 \left({\frac{V_{\rm s}}{16,000 \  {\rm km \ s^{-1}}}}\right)^2 {\left(\frac{t_{\rm trans}}{50 {\rm \ days}}\right)}^2 \ {\rm M}_\odot \ .
\end{eqnarray}

If the wind continues as a $\rho \propto r^{-2}$ wind inside $R_{\rm s}(50 \ {\rm days})$,  the total CSM mass may be considerably higher. The estimate above agrees well with that in \cite{nora}, using a different method. 

\subsection{Nature of SN 2019zrk}

\subsubsection{General properties of the progenitor and ejecta}

As discussed in Sect. \ref{sec:lc}, the powering of the light curve is dominated by CSM interaction, and only a modest mass of ${}^{56}$Ni is allowed. In addition to the spectra discussed earlier, CSM interaction is also demonstrated by the bump in the light curve at $t_{\rm bump} \approx 110$ days in Fig.~\ref{fig:bolom}, where interaction with a strong density enhancement in the CSM is the most natural explanation. If we take the exponential decay, shown as the dotted line in Fig. \ref{fig:bolom} as the reference, the amplitude of the bump in luminosity is $\sim 95$\%, and the area between these curves corresponds to an extra energy radiated in the bump of $\sim 1.8 \times10^{48}$ erg.
With an ejecta velocity of $\sim 16,000 \kms$ (Sect.~\ref{sec:spec_disc}) one can estimate a distance of 
$R_{\rm shell} \approx  1.5 \times 10^{16}$ cm
to such a shell, assuming that it is spherically symmetric. This agrees well with the photospheric radius determined from the blackbody fit in Fig. \ref{fig:teff_rph_luml} at 20 days, which corresponds to a velocity of $\sim 17,500 \kms$. After 20 days the photospheric radius stays nearly constant and therefore recedes in velocity behind the shock.  The agreement between the maximum photospheric  and spectroscopic velocities argues for that the ejecta expansion does not depart dramatically from  spherical. 
If the CSM instead has a ring- or disk-like geometry, or is very clumpy, the velocity and therefore the distance may be overestimated. This was demonstrated for SN 1987A where ejecta with the  highest velocities of $\ga 10^4 \kms$  are expanding above and below the ring plane, while ejecta interacting with the ring have a velocity of $\la 5000 \kms$ \citep{Fransson2013}.  

The duration of the light-curve bump in SN 2019zrk is $\sim 25$ days. If the shock velocity is constant, the thickness of the shell would be $\sim 25/110 \times R_{\rm shell} $. However, because the density in the shell or disk must be higher than the density the shock encountered before the shell interaction, $\rho_0$, the velocity will  be lower by a factor $\sim (\rho_{\rm shell}/\rho_0)^{1/2}$, where $\rho_{\rm shell}$ is the shell density. The above estimate of the thickness is therefore only an upper limit. 

The main conclusion is that there is strong evidence for previous massive ejections of the progenitor. The precursor seen before the main peak of the LC is clear evidence for such an eruption. It is therefore likely that the CSM is more complex than a steady wind and that the light curve and spectra reflect the interaction between different eruptions with different velocities during the decades
before the main eruption in 2020. Although most eruptions of this kind have been discussed in the context of single stars, light-curve bumps have also been discussed as a result of binary interactions \citep{Schwarz1996}. In this case the undulations are expected to be periodic, as may have been the case for SN 1979C \citep{Montes2000}, but for which there is little evidence for SN 2019zrk.

One may  speculate that the bump at $\sim 110$ days is a result of interaction with a shell from a previous eruption. This would have occurred at $t_{\rm bump} V_{\rm s} / V_{\rm shell}$, where $V_{\rm shell}$ is the ejected velocity. $V_{\rm shell}$ is highly uncertain and could range from the estimated CSM velocity, on the order of $100 \kms$, to a velocity similar to what we see in the main eruption, $\ga 10^4 \kms$. 
A velocity of $\sim 13,000 \kms$ was observed for the precursor of SN 2009ip \citep{Smith2010,Pastorello2013,Mauerhan2013}, and assuming a similar velocity for SN 2019zrk, the previous eruption would have occurred $180 (V_{\rm shell}/10^4 \kms)^{-1}$ days before the main event. There is, however, no indication from the pre-explosion light curve (Fig. \ref{fig:ptf_ztf_lc}) of such an event. If we assume a precursor magnitude brighter than the upper limit  of $\sim -15$, which may indicate either a fainter absolute magnitude or a lower $V_{\rm shell}$, increasing the timescale.

The high ejecta velocity seen in the last spectra indicates little deceleration of the shock. This in turn points to a low CSM mass. A low CSM mass is also consistent with the rapid decline of the LC, as well as the relatively low luminosity compared to high-luminosity Type IIn SNe.
This is also consistent with the lower velocities found for high luminosity, long duration objects like SN 2010jl \citep{fransson10jl} and SN 2013L \citep{Taddia2020}, which have velocities  $\sim 5000 \kms$. 
Even at the last epochs we do not see any [O I] $\wll 6300, 6364$ emission in the spectra of SN 2019zrk. This may have have two explanations. First, the processed 
metal-rich core at low velocities  may even at the last epoch be hidden behind the photosphere by optically thick electron scattering. We also note that the velocity corresponding to the photospheric radius at 100 days, $R_{\rm ph} \sim 2\times 10^{15}$ cm, corresponds to a velocity of $\sim 2300 \kms$, so this is not unreasonable.  Alternatively, there is no advanced nucleosyntesis, and the observed transient is not the final core collapse.  Even in the last spectrum the photosphere is far outside the center at $\sim 6000 \kms$, which makes the nucleosynthesis constraint weaker.

Summarizing,
we have a progenitor which has had at least one eruption before the main eruption and probably more as evidenced by the bump in the late LC,  a limited period of very strong mass loss, $\dot M \sim 0.04\  \rm M_{\odot} \ year^{-1}$, and low CSM velocity. There is evidence for a nonuniform CSM, both before the main eruption and in connection to this, indicating several shells from different eruptions. The ejecta have a very high velocity, $\sim 16,000 \kms$, with little deceleration and no obvious indications of advanced nucleosynthesis. 

\subsubsection{Comparison with other SNe}
\label{sec:comparison}
As illustrated in Fig.~\ref{fig:comp_photom}, there  are strong similarities between the light curves of SN 2019zrk and those of SNe 1998S, 2009ip, and 2016bdu, 
both in the rising and in the declining parts. 
SN 1998S was only observed in the declining phase \citep{Fassia2000}, but it shows both a similar decline rate and peak magnitude as SN 2019zrk. Unfortunately, there are no constraints on a precursor phase for this SN. The spectral evolution also has some similarity to SN 2019zrk in that it showed an early symmetric \Ha \ line during the first week, indicating electron scattering and a dense CSM close to the SN  \citep{Leonard2000,Fassia2001,Shivvers2015}.
It also showed strong high-ionization lines of \ion{C}{iii-iv} and \ion{N}{iii}, now referred to as flash-ionization lines \citep{GalYam2014}. Our first high-quality spectrum of SN 2019zrk at 10 days does not show this, but this phase may have been missed. The further evolution of SN 1998S, however, shows important differences to SN 2019zrk both in the evolution of the \Ha \ profile and in the presence of strong absorption and emission lines of \ion{He}{i}, \ion{O}{i}, \ion{Si}{ii}, and \ion{Fe}{ii}. Although SN 1998S showed strong CSM interaction from the \Ha \  line and the light curve, it therefore differs markedly from SN 2019zrk. 
 
SN 2009ip is one of the best studied SNe dominated by CSM interaction  
\cite[see][for a review]{Fraser2020}. 
For SN 2009ip the sharp rise in the main outburst on 25 September 2012 occurred on a timescale of about a day from $M_R \approx -14.5$ to $M_R \approx -18$ \citep{Prieto2013,Pastorello2013,Mauerhan2013}. This is faster than for SN 2019zrk, where the rise from $M_r \approx -16.2$ to $M_r \approx -17.2$ took place in $\sim 6$ days. For SN 2019zrk the rise then accelerated and the final rise to $M_r \approx -18.6$ took place in just two days (Fig. \ref{fig:comp_photom}). The declining part of the light curve is slower for SN 2019zrk, but not by a large factor. The  main difference is in the peak magnitude, where SN 2019zrk was $\sim 1$ magnitude brighter than both SN 2009ip and SN 2016bdu \citep{Pastorello2018}.

The light curve bump at $\sim 110$ days is also not unique to SN 2019zrk. In SN 2009ip a temporary brightening occurred already at $\sim 42$ days after the final rise by $\sim 0.3$ mag, lasting a few days \citep{Graham2014}. While the timescale is different, they are in both cases strong indications of a nonuniform, dense CSM.
Another important similarity is the presence of a precursor in both SNe 2009ip and 2019zrk. 
Prior to the main outburst in 2012, SN 2009ip had two strong outbursts, to $M_R \approx -14 $,  the first $\sim 1100$ days  before and the second $\sim 500$ days before the main outburst \citep{Pastorello2013, Mauerhan2013}. These lasted 100--200 days, and showed strong fluctuations between $M_R \sim -11$ to $M_R \sim -14$ on timescales of a few days. 
In contrast to SN 2019zrk, SN 2009ip 
had spectroscopic coverage also before the most energetic event. 

SN 2009ip could be followed photometrically and spectroscopically for more than 1000 days \citep{Fraser2015,Graham2017}. Of special interest is the fact that the bolometric light curve at late epochs, up to $\sim 700$ days, was flatter than that corresponding even to full trapping of ${}^{56}$Co, with an upper limit of a ${}^{56}$Ni mass of $\la 0.02$ \msun~ \citep{Fraser2015,Graham2017}. The luminosity of a radiative reverse shock is expected to follow $L \propto \dot M V_{\rm s}^3/u_{\rm w} \propto t^{-3/(n-3)}$, where $n$ is the power law of the ejecta density profile. With $n \approx 10$, this indicates that CSM interaction still dominated the light curve of SN 2009ip at these very late phases. 

\begin{figure}
\centering
\includegraphics[width=9cm,angle=0]{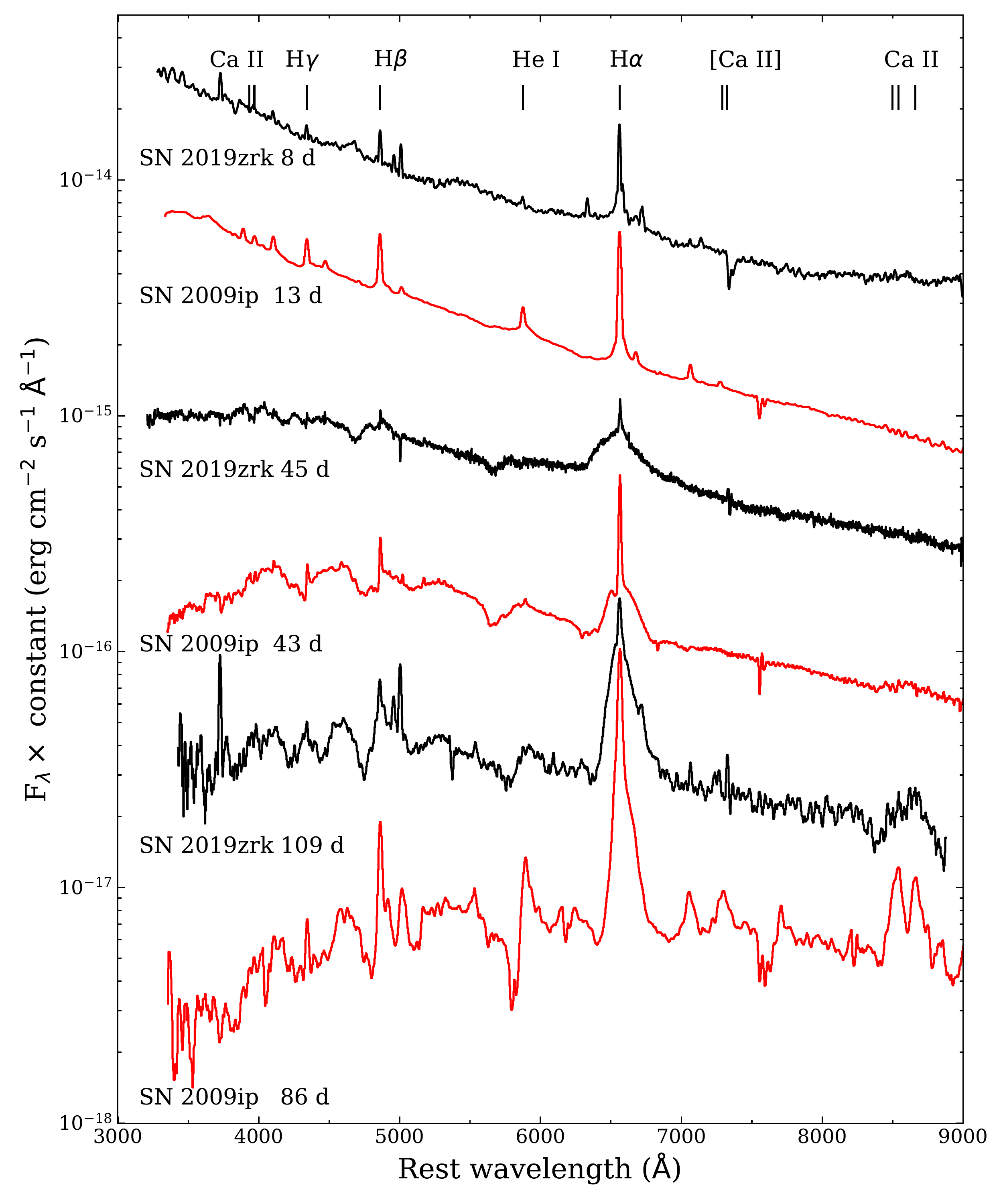}  
\caption{\label{fig:comp_19zrk_09ipl} Comparison of the spectral evolution of SN 2019zrk and SN 2009ip (red) from \cite{Fraser2013} at three comparable epochs.  
}
\end{figure}

The spectral evolution of SN 2019zrk also show strong similarities with that of SN 2009ip. Several groups have presented extensive series of spectra both before and after the main outburst for that SN \citep{Smith2010,Pastorello2013, Mauerhan2013,Fraser2013,Margutti2014,Graham2014}, in particular 
presenting a very complete set of observations up to $\sim 100$ days, which directly can be compared to SN 2019zrk. In Fig. \ref{fig:comp_19zrk_09ipl} we show the spectra of SN 2019zrk at three representative epochs together with spectra at similar stages for SN 2009ip, taken from \cite{Fraser2013}.

Immediately after 
the main eruption of SN 2009ip
a symmetric 'exponential' emission line profile, typical of electron scattering, was observed, resembling that seen in the spectra of SN 2019zrk during the first 21 days.  In SN 2009ip this line profile persisted until 
$\sim25$ days after the eruption. After $\sim45$ days \Ha \ had become asymmetric with a P-Cygni absorption at high velocity and a broad emission profile. H$\beta$ developed a broad P-Cygni profile already at $\sim25$ days. The absorption part of this line profile extended to $\sim 15,000 \kms$, while the emission extended only to $\sim 10,000 \kms$. Also the He I $\wll 4471, 5876$ lines had strong absorption components and narrow, weak emission.  The spectrum of SN 2009ip from  $\sim 90$ days
showed a strong, asymmetric \Ha \ line, while other Balmer and He I  lines are primarily seen in absorption. These spectra strongly resembles our last spectrum of SN 2019zrk, although the evolution of the SN 2009ip spectrum is somewhat faster and the velocities lower, as can for example be seen in the Ca II triplet, which is resolved in SN 2009ip at 86 days.  All in all, however, the spectral evolution of SN 2019zrk is very similar to that of SN 2009ip.

\cite{Levesque2014} discussed the Balmer ratio for SN 2009ip  finding an \Ha/\Hb \ ratio  $\sim 1.5$ at the peak of the main eruption, indicating an electron density of $\ga 10^{13} \ \ccm$. We estimate a ratio of $\sim 2.5$ in our 
spectrum at 10 days, but the value is uncertain because of the sensitivity to the subtraction of the continuum and the influence of the background  
contamination. In the later epochs the ratio increases (Sect. \ref{sec:spec}) similar to what is the case for the Type IIn SN 2010jl \citep{fransson10jl}. The evolution in SN 2019zrk is consistent with these two SNe, but because of the strong  P-Cygni absorption in \Hb \ in SN 2019zrk a more quantitative comparison is difficult.

The main difference compared to SN 2009ip and to SN 2016bdu is the higher peak luminosity of SN 2019zrk. This could perhaps be connected to the higher precursor luminosity of SN 2019zrk, which may indicate a more dense CSM into which the ejecta expand, resulting in a higher shock luminosity. 
Unfortunately, we do not have any spectra of the precursor, but we just note the remarkable high velocities of the precursor of SN 2009ip, up to $\sim 13,000 \kms$ \citep{Smith2010,Pastorello2013,Mauerhan2013}, and the interesting spectral evolution of that SN during the two preceding eruptions. 

In the case of SN 2009ip  very late spectra and photometry up to $\sim 1000$ days have been presented by \cite{Fraser2015} and \cite{Graham2019}. No dramatic changes in the spectrum from the spectra at $\sim 100$ days are seen. In particular, the Balmer lines remain symmetric. Even at these late stages the [\ion{O}{i} $\wll 6300, 6364$ doublet remains weak in contrast to the late evolution of Type IIP SNe. Graham et al. also find that the \Ha \ and photometric evolution agree well with the predictions in \cite{CF94}, supporting the dominance of CSM interaction. 

After
SN 2009ip several other Type IIn SNe have been discovered with very similar properties, including SN 2010mc \citep{Ofek2013,Smith2014a}, SN 2016bdu \citep{Pastorello2018}, and  AT 2016jbu = Gaia16cfr \citep[][]{Kilpatrick2018,Brennan2021,Brennan2021b}.

SN 2010mc, which besides SN 2009ip was the first Type IIn which showed a clear precursor, showed many similarities to SN 2009ip \citep{Ofek2013,Smith2014a}, as well as SN 2019zrk. The length of the precursor, $\sim 50$ days, was somewhat shorter than or SN 2019zrk, while the rise time, $\sim 7$ days was similar. The peak magnitude of the precursor, M$_{\rm r} \approx -15$ as well as the main event,  M$_{\rm r} \approx -18.4$, were, however, approximately one magnitude fainter. Consequently, the total enery radiated by both the precursor and the main event were lower by a actor $2-3$. 

\cite{Kilpatrick2018} and \cite{Brennan2021,Brennan2021b}  present a nice set of spectral and photometric observations for AT 2016jbu, which 
shows a precursor with a similar duration as for SN 2019zrk, $\sim 100$ days, and with a similar gradual increase in magnitude. The main event showed a similar photometric evolution to SN 2009ip, both in absolute magnitude and in the light curve shape. 
Compared to SN 2009ip and SN 2019zrk, no bump in the light curve was, however, seen during the first $\sim 100$ days after the main event in AT 2016jbu. 
Instead, it 
showed a strong increase in the luminosity at $\sim 220$ days. 

Although the light curve was similar, the spectral evolution of AT 2016jbu was quite different from that of SNe 2019zrk and 2009ip. In addition to an \Ha \ emission line peaking at zero velocity, AT 2016jbu also showed a double peaked profile with a blue emission peak at $\sim -3000 \kms$. This blue emission emerges at $\sim +20$ days after peak luminosity, becoming equally strong as the zero velocity peak at $+ 50$ days and later. 
 
 SN 2016bdu also had a remarkably similar evolution to SNe 2019zrk and 2009ip, exhibiting a precursor, a fast rise and a steep  decline of the main event 
 (Fig.~\ref{fig:comp_photom}). 
Like SN 2009ip it showed a flat late phase LC $\sim 100$ days after the main eruption. The length of the precursor was $\sim 70$ days, similar to that for SN 2019zrk. The progenitor also showed an erratic behavior over longer timescales
with outbursts to $M_R\sim -14.5$.

\cite{Pastorello2018} also point out the similarity of the properties of the 09ip-like SNe to SN 2005gl, for which \cite{Galyam2007} found evidence for a massive LBV-like progenitor, which was found to have disappeared after the decline of the SN \citep{Gal-Yam2009a}. Both the spectral evolution and the light curve of SN 2005gl show strong similarities to SN 2009ip, with an only marginally fainter peak and similar rise and decay timescales. In contrast to SN 2009ip and SN 2019zrk, SN 2005gl had, however, no detectable precursor down to M $\approx -13.7$.

Most recently, \cite{Pessi2022} present monitoring of SN 2011fh from 2007 to 2017. Before the main eruption the star stayed at a steady level with M$_r \sim -14.5$ for at least three years. This was followed by a precursor with a duration of $\sim 200$ days with M$_r \sim -16$ culminating in the main eruption. The peak magnitude was M$_r \sim -17.9$. The decline was initially fast but slowed down after $\ga 100$ days, and was still bright 1300 days after the peak at M$_r \sim -13.5$, similar to the peak brightness of Eta Carinae
From HST photometry of the stellar cluster around SN 2011fh they find an age of $\sim 4.5$ Myr, corresponding to the life time of a star $35 -80 \Msun$, depending on being in a binary or single. This assumes a single burst with zero spread in age, which is an important caveat.

One may also compare SN 2019zrk with other related classes of interacting SNe, in particular with 
Type Ibn SNe and transition types like SN 2014C.
Using the Type Ibn light-curve template by \cite{Hosseinzadeh2017} we find that the rate of decline of SN 2019zrk is a factor of $\sim 2$ slower than for this class of SNe. This may be a result of a difference in opacity between a helium and hydrogen dominated CSM, but could also reflect different masses of the CSM. 

Possibly related SNe are SN 2001em and SN 2014C, which represent transitions from a Type Ib SN to a more Type IIn like spectrum, in parallel with an increase of the \Ha, radio and X-ray luminosity \citep{Chugai2006,Chandra2020,Milisavljevic2015,Margutti2017}. For SN 2001em this transition occurred $\sim 2$ years after explosion and for SN 2014C at $\sim 100$ days. Both the observations and modeling \citep{Chugai2006} pointed to interaction with a previously ejected circumstellar shell at a distance of $\sim (5-7)  \times 10^{16}$ cm. Although the light curves and spectral evolution are very different from those of SN 2019zrk, the presence of a shell, as we infer from the bump at 110 days, indicates strong similarities. A possibility is that these SNe represent a more advanced stage of evolution where the hydrogen envelope has been completely ejected during previous outbursts. 

\subsection{Origin of the CSM and nature of the progenitor}
\label{sec:Origin}

Summarizing the last section, the 09ip-class of SNe, including SN 2019zrk, show a remarkable similarity in light curves, spectra and energetics, although there are factor of two differences in these properties between the different object, with SN 2019zrk being among the most energetic. It is therefore based on these results interesting to discuss different progenitor scenarios.

For SN 2009ip there has been a long debate about whether the main eruption was 
due to the final core collapse, or only a particularly violent eruption. This relates to whether the progenitor had a very large mass, similar to that of pulsational pair instability SNe (PPISN), $\ga 80 \ \Msun$, or if the progenitor was 
in the lower mass range, $8 - 20 \ \Msun$.  
Connected to this, there is a need to explain the existence of a very dense CSM close to the progenitor. 

An argument against a lower mass core-collapse SN may be absence of significant ${}^{56}$Ni production, as indicated from the tail of the light curve, or possibly the lack of indications of advanced nucleosynthesis from the late spectra. This would instead be an argument in favor of a PPISN.  For SN\,2009ip, \cite{Fraser2015} find an upper limit of $M({}^{56}$Ni)  $\la 0.02 \ \Msun$, while \cite{Brennan2021} find $M({}^{56}$Ni) $\la 0.016 \ \Msun$ for AT 2016jbu. For SN 2019zrk we only have an upper limit of $\la 0.09 \ \Msun$. A core collapse, with a moderate  $M({}^{56}$Ni mass, can therefore not be ruled out.

Observations by \cite{Thoene2015} showed that by 
late 2015, SN 2009ip had declined to a magnitude below that before the 2012 eruption. The absence of a detected star
could argue for a core collapse scenario. However, as remarked by \cite{Graham2017}, the progenitor could have a fainter absolute magnitude in a quiescent phase, so this observation is not conclusive.

In the case of detections of progenitors there is also a debate of the inferred masses of these. 
In the case of SN 2009ip, the detection with the {\it Hubble Space Telescope (HST)} at $M_V \approx -9.8$ approximately 10 years before the 2009 eruption, may correspond to a ZAMS mass of $50-80 \ \Msun$ \citep{Smith2010}, although \cite{Foley2011} only declared a lower limit of $\ga 60$ \msun. 
This mass determination assumes that the star was in a quiescent state at the time of the 1999 observation, and 
the determination is only based on a single band, the F606W.  The high progenitor mass argued for SN 2009ip is close to that of 
PPISNe, although at the lower limit of these $\sim 80$ \msun \  \citep[e.g.,][]{Woosley2017}. 
A PPISN scenario would also explain the large CSM mass observed.

The high mass range for the progenitors of the 09ip-class of SNe has been questioned by \cite{Kilpatrick2018} and \cite{Brennan2021b} in relation to AT 2016jbu. For this SN, multiband
 {\it HST} photometry of the progenitor $\sim 10$ years before the main event indicate a yellow hypergiant with mass $17-22 \ \Msun$ \citep{Brennan2021b}, which is also consistent with the local stellar population in the neighborhood. \cite{Kilpatrick2018} point out that the progenitor was probably reddened, and the mass may have been up to $\sim 30 \ \Msun$. They also point out that the strong contamination of  \Ha \ in bands covering this wavelength can lead to a large overestimate of the progenitor mass. This is the case for the F606W filter with WFPC2, which may explain the discrepant masses for SN 2009ip and AT 2016jbu. 
The high host-galaxy metallicity (Sect. \ref{sec:host}) for SN 2019zrk also argues against a PPISN.

Another argument  against an origin from a PPISN is the very high ejecta velocity seen in the last spectra of SN 2019zrk. Models of PPISNe typically result in velocities of $\la 5000 \kms$ \citep{Woosley2017}. We note here, however, that for the LBV Eta Carinae, which may be in a PPI phase \citep{Woosley2017},  material with velocities of $\sim 6000 \kms$ \citep{Smith2008}, and possibly up to $20,000 \kms$ \ \citep{Smith2018} have been observed, although the main shell from the year 1842 eruption is moving with only $600 \kms$ \citep{Smith2018}. This indicates several eruptions with highly different velocities, or an asymmetric eruption. In the end there may thus be several problems with the PPISN scenario for the 09ip-class, and we find this less likely. 

A possibility for creating a dense CSM in stars with lower masses is by wave heating, powered by gravity waves or nuclear flashes in the core in connection to the last burning stages before core collapse \citep{Shiode2012,Woosley2015,Fuller2017,Wu2021,Linial2021}. The total energy in these waves as they reach the hydrogen envelope is typically $ 10^{46} - 10^{47}$ erg \citep{Wu2021}, which is hardly enough to unbind the H envelope.  \cite{Wu2021}, however, find that for low-mass progenitors, $\la 12 \ \Msun$, and for massive progenitors, $\ga 30 \ \Msun$, there may be cases where the energy is larger and could cause mass ejections.
For the low masses this occurs $0.1 - 10$ years before core collapse during O/Ne burning, while in the higher mass range this occurs as a result of shell mergers, only $0.01-0.1$ years before core collapse. For an extensive H envelope this timescale may be too short to propagate to the surface. However, for stripped SNe it may have important consequences.    

In this connection we also note that observations of 'flash ionization' features during the first days after explosion indicate that a large fraction of Type II SNe, $\ga 30 \%$, have a very dense CSM \citep{Bruch2021}, not compatible with ordinary stellar winds. The masses discussed there are, however, considerably lower than for the 09ip-class. This does show that our understanding of the mass loss in the last stages of massive stars is far from complete. 

The low limits for the ${}^{56}$Ni production, discussed above, may be a problem for stars with ZAMS masses $\ga 12 \ \Msun$, while lower masses only produce smaller amounts of ${}^{56}$Ni, compatible with the limits for the 09ip-class.
\cite{Brennan2021b} and \cite{Pastorello2019} also discuss luminous red novae, proposed as a result of mergers between two massive stars as a possible channel for the 09ip-class of SNe. As \cite{Brennan2021b} point out, the peak magnitudes of these mergers are, however, $\sim 3$ mags fainter than the peak magnitude of SN 2009ip and AT 2016jbu, and $\sim 4$ mag fainter compared to SN 2019zrk.  

A related possibility is the merger between a massive star and a compact object. This was discussed by 
\cite{Chevalier2012} in the context of Type IIn SNe and was also discussed as a possibility for SN 2001em \citep{Chandra2020}. Loss of angular momentum leads to an outflow in the orbital plane and a dense circumstellar disk. At the same time accretion onto the compact object liberates energy, mainly in the form of neutrinos, but also as electromagnetic radiation and Poynting flux. The interaction of the blast wave created by the compact object and the disk then gives rise to the optical display and light curve. The accretion rate depends on the angular momentum of the gas, which depends on the density and velocity gradient of the companion \citep{Chevalier1993}. \cite{MacLeod2015} argue that this may prevent the neutron star from collapsing to a black hole, although hypercritical accretion may still occur. In either case the accretion energy may power the explosion \citep{Fryer1998}. 

\cite{Schroder2020} have made detailed simulations of such a merger and in particular studied the common-envelope structure for different parameters. In the inspiral of the companion $\sim 25 \%$ of the mass of the compact object is expelled from the donor star, or $4.5 - 13.9 \ \Msun$, for a mass ratio of the compact object and donor between 0.1 -- 0.3. The energy released from accretion onto the compact object is assumed to be deposited roughly spherically in the hydrogen envelope. In general they find a density profile which is fairly steep with radius, like $\rho \propto r^{-3}$ or $\rho \propto r^{-4}$, depending on the mass ratio of the compact object and the primary star.
Using the SNEC code \citep{Morozova2015} they calculate the light curve resulting from the interaction of such a SN with the dense CSM. The results are compared to the Type IIL SNe 1979C and 1998S, and in both cases the general photometric evolution is well reproduced. This scenario has several appealing properties in being able to provide very large luminosities from the orbital energy, reasonable rates, and providing a dense CSM just before the explosion.  As shown in Fig. \ref{fig:comp_photom}, the LCs of SN 1998S and SN 2019zrk are  very similar, adding to this interpretation, although the existence of  a precursor for SN 1998S is not known. 

 \cite{Terman1995} argue that binary systems with orbital periods less than $0.2 - 2$ years for companion masses of $12 - 24 \ \Msun$ will not be able to eject the common envelope and instead undergo a complete merger. This agrees well with more recent simulations based on MESA models for the companions by \cite{MacLeod2015}. Based on this, and that the timescale of the merger should be of the order of the orbital period, one may speculate that the precursor seen in the 09ip class of objects corresponds to the energy release during the final inspiral of the merger. 
 
A possible problem for the merger model may be the large velocity seen in the precursor of SN 2009ip, $\sim 12,000 \kms$ \citep{Foley2011,Pastorello2013}. In the simplest scenario one might expect a velocity similar to the escape velocity of the core,  $V_{\rm escape} = 364 \ (M_{\rm core}/5 \ \Msun)^{1/2} (R_{\rm core}/10^{12} \ {\rm cm})^{-1/2} \  \kms$. However, no detailed modeling of the effect of the energy release during the inspiral of a neutron star into the envelope and core has been made.
In addition, the fact that Eta Carinae showed ejecta velocities at least up to $6000 \ \kms$ in connection to the 1843 eruption \citep{Smith2008} shows that we do not understand the details of the final stages of massive stars.

An interesting observation is the dip in the light curve between the peak of the precursor and the rise of the main event, seen for SN 2009ip \citep{Pastorello2013,Mauerhan2013}, SN 2010mc \citep{Ofek2013}, SN 2016bdu \citep{Pastorello2018}, AT 2016jbu \citep{Brennan2021} and SN 2019zrk (Fig. \ref{fig:comp_photom}). A similar dip is seen for the best observed merger in the Milky Way, V1309 Sco \citep[Fig. 1 in ][]{Tylenda2011}, although on a different timescale and luminosity, which may be a result of the highly different masses and separation of the systems. If this dip is generic to the merger scenario it may offer an interesting diagnostic and clue to the progenitor. Another feature of the light curve of V1309 Sco is the bump in the declining part of the light curve, seen also for SN 2009ip and SN 2019zrk.

Summarizing this discussion, there are potential issues with all the proposed scenarios for the 09ip-class of SNe, although we believe that the low mass alternative, involving a merger, is the one with the least problems. 
The limited theoretical understanding of the complicated interplay between nuclear burning and hydrodynamics during the last nuclear burning stages, as well as the merger scenario, in particular the common envelope phase, make conclusions based on modeling uncertain for the whole mass range discussed.

\section{Summary and conclusions}\label{sec:conclusions}

From the spectral and photometric evolution, we have shown that SN 2019zrk displays many properties in common with the 09ip-class of SNe. The presence of a precursor, the sharp LC rise, the fast, exponential LC decay, and the LC bumps are all very similar, as is the spectral evolution. The main, minor difference is the brighter peak magnitude of SN\,2019zrk, as well as the brighter precursor. The total radiated energy in the main event is $\ga 5\times 10^{49}$ erg, and in the precursor $\ga 4\times 10^{48}$ erg. The spectral evolution from exponential electron scattering emission line profiles to broad P-Cygni lines are clear indications of circumstellar interaction. The extent of the P-Cygni absorption indicates a high ejecta velocity of $\sim 16,000 \kms$. From the time of optical depth unity to electron scattering, we estimate a high mass-loss rate of $\sim 4.4 \times 10^{-2} (u_{\rm w}/100 \ \kms)  \  \rm M_{\odot} \ yr^{-1}$. The total mass in the CSM is $\ga 1 \ $ \msun. In common with SN 2009ip, it shows a clear bump in the light curve, which may be the result of interaction with circumstellar material from a previous eruption, for example similar to the one seen as a precursor. 

The origin of the 09ip-class of SNe is still unclear. The absence of products of advanced nucleosynthesis and the fast light curves are compatible with a 
pulsational pair instability SN. What argues against this scenario is the very fast ejecta velocities observed for both SNe 2009ip and 2019zrk, and also the low mass inferred for the progenitor of AT 2016jbu. The rate of this type of SNe may also pose a problem for a large ZAMS mass progenitor model, although good statistics of the 09ip-class of SNe is lacking.  An origin in a lower-mass progenitor, $\la 20$ \Msun, with a
mass ejection during the last phases before core collapse may have fewer problems, although the mechanism of these large mass eruptions is not well understood.  We propose the merger scenario as the most promising for explaining both the dense CSM, seen in the precursor and main eruption, and the large peak luminosity. Clearly both more observations of this type of SNe are needed, as well as a better theoretical understanding of especially mass loss processes close to core collapse.

\begin{acknowledgements}
We are grateful to Jacob Nordin and the referee for comments and careful readings of the paper. 
The research of C.F. is supported by the Swedish Research Council. NLS is funded by the Deutsche Forschungsgemeinschaft (DFG, German Research Foundation) via the Walter Benjamin program – 461903330. S.S. and E.C.K. acknowledges support from the G.R.E.A.T research environment, funded by {\em Vetenskapsr\aa det},  the Swedish Research Council, project number 2016-06012, as well as support from the Wenner-Gren Foundations.  We thank Jakob Nordin for comments on the manuscript.
Based on observations obtained with the Samuel Oschin Telescope 48-inch and the 60-inch Telescope at the Palomar Observatory as part of the Zwicky Transient Facility project. 
ZTF is supported by the National Science Foundation under Grant No. AST-1440341 and a collaboration including Caltech, IPAC, the Weizmann Institute for Science, the Oskar Klein Center at Stockholm University, the University of Maryland, the University of Washington, Deutsches Elektronen- Synchrotron and Humboldt University, Los Alamos National Laboratories, the TANGO Consortium of Taiwan, the University of Wisconsin at Milwaukee, and Lawrence Berkeley National Laboratories. Operations are conducted by COO, IPAC, and UW. 
This work was supported by the GROWTH project funded by the National Science Foundation under PIRE Grant No 1545949. 
The Oskar Klein Centre was funded by the Swedish Research Council.
Partially based on observations made with the Nordic Optical Telescope, 
 owned in collaboration by the University of Turku and Aarhus University, and operated jointly by Aarhus University, the University of Turku and the University of Oslo, representing Denmark, Finland and Norway, the University of Iceland and Stockholm University at the Observatorio del Roque de los Muchachos, La Palma, Spain, of the Instituto de Astrofisica de Canarias. 
Some of the data presented here were obtained with ALFOSC. 
Some of the data presented herein were obtained at the W. M. Keck
Observatory, which is operated as a scientific partnership among the
California Institute of Technology, the University of California, and
NASA; the observatory was made possible by the generous financial
support of the W. M. Keck Foundation. 
The Liverpool Telescope is operated on the island of La Palma by Liverpool John Moores University in the Spanish Observatorio del Roque de los Muchachos of the Instituto de Astrofisica de Canarias with financial support from the UK Science and Technology Facilities Council.
The SED Machine is based upon work supported by the National Science Foundation under Grant No. 1106171. 
\end{acknowledgements}

\clearpage

\bibliography{sn2019zrk}

\clearpage

\clearpage

\begin{table*}
\caption{Summary of photometric observations. 
The full table will be made available in machine-readable format.
} 
\label{tab:phot}
\centering 
\begin{tabular}{lccccccl}
\hline\hline
Date & Filter & Phase & Magnitude & Magnitude Error & limiting mag & Absolute mag & Instrument \\
(JD) & & (Rest-frame days) & & & & & \\
\hline

\hline
2458784.04 & r  & -96.49 & 20.82 &  0.35 & 20.30 & -15.19 & P48+ZTF\\ 
2458786.02 & r  & -94.58 & 99.00 & 99.00 & 20.19 & -15.77 & P48+ZTF\\ 
2458790.02 & g  & -90.72 & 99.00 & 99.00 & 19.68 & -16.28 & P48+ZTF\\ 
2458793.05 & g  & -87.79 & 99.00 & 99.00 & 20.11 & -15.85 & P48+ZTF\\ 
2458803.06 & g  & -78.13 & 99.00 & 99.00 & 19.85 & -16.11 & P48+ZTF\\ 
2458806.02 & g  & -75.27 & 99.00 & 99.00 & 20.06 & -15.90 & P48+ZTF\\ 
2458829.04 & g  & -53.05 & 99.00 & 99.00 & 19.77 & -16.19 & P48+ZTF\\ 
2458829.07 & r  & -53.02 & 20.29 &  0.18 & 20.48 & -15.72 & P48+ZTF\\ 
2458831.96 & r  & -50.24 & 19.68 &  0.31 & 19.30 & -16.34 & P48+ZTF\\ 
2458833.98 & r  & -48.28 & 99.00 & 99.00 & 19.74 & -16.22 & P48+ZTF\\ 
2458837.96 & r  & -44.44 & 20.00 &  0.30 & 19.63 & -16.01 & P48+ZTF\\ 
2458838.02 & g  & -44.38 & 20.29 &  0.24 & 20.19 & -15.75 & P48+ZTF\\ 
2458839.02 & r  & -43.42 & 99.00 & 99.00 & 19.49 & -16.47 & P48+ZTF\\ 
2458846.96 & r  & -35.75 & 20.18 &  0.14 & 20.69 & -15.83 & P48+ZTF\\ 
\hline
\end{tabular}
\end{table*}

\begin{table*}
\caption{Summary of spectroscopic observations \label{tab:spec}}
\begin{tabular}{lcclc}
\hline\hline
Object&Observation Date&Phase&Telescope+Instrument&Resolution \\
 &&(Rest-frame days)&&{($\kms$)}\\
\hline
SN\,2019zrk & 2020 Feb 13 & \phantom{11}8.4 & P60+SEDM&3000\\
SN\,2019zrk & 2020 Feb 14 & \phantom{1}10.0 & P200+DBSP&\phantom{1}130\\
SN\,2019zrk & 2020 Feb 15 & \phantom{1}11.4 & P60+SEDM&3000\\
SN\,2019zrk & 2020 Feb 17 & \phantom{1}12.2 & LT+SPRAT&\phantom{1}830 \\
SN\,2019zrk & 2020 Feb 17 & \phantom{1}12.4 & APO+DIS&\phantom{1}370 \\
SN\,2019zrk & 2020 Feb 24 & \phantom{1}20.0  & P60+SEDM&3000 \\
SN\,2019zrk & 2020 Feb 25 & \phantom{1}20.8 & NOT+ALFOSC&\phantom{1}640 \\
SN\,2019zrk & 2020 Mar 03 & \phantom{1}26.9  & P60+SEDM&3000 \\
SN\,2019zrk & 2020 Mar 22 & \phantom{1}45.3 & Keck+LRIS&\phantom{1}320 \\
SN\,2019zrk & 2020 Mar 28 & \phantom{1}51.1 & P60+SEDM&3000 \\
SN\,2019zrk & 2020 Apr 17 & \phantom{1}70.2 & P60+SEDM&3000 \\
SN\,2019zrk & 2020 Apr 29 & \phantom{1}82.5 & NOT+ALFOSC&\phantom{1}640 \\
SN\,2019zrk & 2020 May 26 & 107.8 & P60+SEDM&3000 \\
SN\,2019zrk & 2020 May 27 & 109.5 & NOT+ALFOSC&\phantom{1}640 \\
\hline
\end{tabular}
\end{table*}

\begin{table*}
\caption{Emission-line measurements of the host galaxy of SN 2019zrk
\label{tab:host_eml}
}
\small
\begin{tabular}{lcc}
\toprule
     & \textbf{Region 1} & \textbf{Region 2}\\
Line & Flux     & Flux\\
     & $\left(10^{-17}\,\rm erg\,cm^{-2}\,s^{-1}\right)$& $\left(10^{-17}\,\rm erg\,cm^{-2}\,s^{-1}\right)$\\
\midrule
H$\beta$                        & $185.02 \pm 5.43$ &$158.00 \pm 2.28$\\
{[}O\textsc{iii}{]}$\lambda$5007    & $103.30 \pm 3.85$ &$ 59.68 \pm 1.69$\\
H$\alpha$                       & $785.54 \pm 4.10$ &$623.82 \pm 2.71$\\
{[}N\textsc{ii}{]} $\lambda$6585     & $247.00 \pm 3.43$ &$204.27 \pm 1.74$\\
\bottomrule
\end{tabular}
\tablefoot{All measurements are corrected for Milky Way extinction. We did not attempt to correct the measurements for stellar absorption. The coordinates of region 1 and 2 are R.A., Decl. (J2000) = 11:39:47.03, +19:55:58.10 and 11:39:47.41, +19:55:53.24, respectively.}
\end{table*}

\begin{table*}
\caption{Photometry of the host galaxy of SN 2019zrk \label{tab:host_phot}}
\small
\begin{tabular}{lcclcc}
\toprule
Survey &  Filter & Brightness   & Survey    & Filter    & Brightness\\
       &         & (mag)        &           &           & (mag) \\
\midrule
\textit{GALEX}  &$ FUV $&$16.14 \pm 0.01$ & PS1             &$ i   $&$13.34 \pm 0.01$\\
\textit{GALEX}  &$ NUV $&$15.64 \pm 0.01$ & PS1             &$ z   $&$13.18 \pm 0.02$\\
SDSS            &$ u'  $&$14.90 \pm 0.01$ & PS1             &$ y   $&$13.02 \pm 0.02$\\
SDSS            &$ g'  $&$13.89 \pm 0.01$ & 2MASS           &$ J   $&$12.87 \pm 0.03$\\
SDSS            &$ r'  $&$13.53 \pm 0.01$ & 2MASS           &$ H   $&$12.88 \pm 0.04$\\
SDSS            &$ i'  $&$13.28 \pm 0.01$ & 2MASS           &$ K   $&$12.78 \pm 0.04$\\
SDSS            &$ z'  $&$13.08 \pm 0.03$ & \textit{WISE}   &$ W1  $&$13.42 \pm 0.01$\\
PS1             &$ g   $&$13.87 \pm 0.01$ & \textit{WISE}   &$ W2  $&$13.88 \pm 0.01$\\
PS1             &$ r   $&$13.45 \pm 0.01$ & \\
\bottomrule
\end{tabular}
\tablefoot{All magnitudes are reported in the AB system and are not corrected for extinction.}
\end{table*}

\end{document}